\documentclass[preprintnumbers]{revtex4}
\usepackage{amssymb}
\usepackage{graphicx}
\usepackage{color}

\def\bc{\begin{center}}
\def\ec{\end{center}}

\def\beq{\begin{equation}}
\def\eeq{\end{equation}}

\input{tcilatex}

\begin{document}

\title{Trions and biexcitons in a nanowire}
\author{R. Ya. Kezerashvili$^{1}$, \ Z. S. Machavariani$^{2}$, B. Beradze$%
^{2}$, and T. Tchelidze$^{2}$}
\affiliation{$^{1}$Physics Department, New York City College of Technology, The City
University of New York, Brooklyn, New York 11201, USA\\
$^{2}$Department of Exact and Natural Sciences, Tbilisi State University,
0179 Tbilisi, Georgia}
\date{\today}

\begin{abstract}
A theory of the trion and biexciton in a nanowire (NW) in the framework of
the effective-mass model using the Born-Oppenheimer approximation is
presented. We consider the formation of trions and biexcitons under the
action of both the lateral confinement and the localization potential. The
analytical expressions for the binding energy and eigenfunctions of the
trion and biexciton are obtained and expressed by means of matrix elements
of the effective one-dimensional cusp-type Coulomb potentials whose
parameters are determined self-consistently by employing the same
eigenfunctions of the confined electron and hole states. Our calculations
for the ZnO/ZnMgO, CdSe/ZnS and CdSe/CdS core/shell cylindrical shaped NWs
show that the trion and biexciton binding energy in NWs are size-dependent
and for the same input parameters the biexciton binding energy in NWs is
always larger than the binding energy of the trion. The trion and biexciton
remain stable in CdSe/ZnS NW with the increase of the dielectric shell,
while in ZnO/ZnMgO NW they become unstable when the surrounding dielectric
shell exceeds 2.5 nm and 2 nm for each, respectively. The associative
ionization of biexciton antibonding states into trion bonding states that
leads to the formation of trions is studied. Based on the results for size
dependence of biexciton binding energy and probability associative
ionization an optimal radius for optoelectronic application NW is suggested.
\end{abstract}

\maketitle


\section{Introduction}

\label{Introduction} 
The optical properties of quantum nanostructures have been increasingly
investigated over the past decades. This is connected to the fact that the
system characteristics that govern optical response, such as electronic
level structure, oscillator strength, Coulomb interaction between charged
carriers and electron-phonon interaction dramatically changes with size
variation at nanometer scale \cite{Science2013,Nanophotonics}. Consequently,
the issue of tuning of optical response by means of size and shape control
has become important. Excitons and excitonic complexes in quantum
nanostructures have been one of the hot topics since the early days of
quantum nanostructures \cite{Efros}. The reason for this is that on the one
hand excitons are main intrinsic emitters in short wavelength region and
therefore, optimization of excitonic emission is very important for emitting
device fabrication. On the other hand, the investigation of excitons and
their complexes can provide deep insight into the peculiarities of inter
particle interaction at low dimensions. Multi-particle states, like single
particle ones, are strongly affected by space and dielectric confinement 
\cite{Giblin}. Therefore, the new possibilities of controlling their
characteristics appear in quantum nanostructures.

In the late 1950s Lampert \cite{L} predicted the existence of charged and
neutral exciton complexes formed when an electron in a conduction band or a
hole in a valence band is bound to a neutral exciton or two single excitons
are correlated. This idea gave rise to many publications in the 60s and the
70s in bulk materials (see, for example, the works \cite{Stebe1974,
Stebe1977, Stebe1987} and citations therein). The binding energies of the
exciton complexes are very small in bulk at room temperature, but they are
substantially enhanced in structures of reduced dimensionality. Theoretical
calculations performed at the end of the 1980s \cite{Stebe1989} predicted a
considerable (up to tenfold) increase of the trion binding energy in quantum
well heterostructures compared with bulk. Trions were first observed in
quantum wells (QW) \cite{QW1} in 1993 and shortly thereafter in GaAs-AlGaAs
quantum wells \cite{QW2,QW3,QW4}.

In the last two decades these complexes have been the subject of an
extensive theoretical and experimental studies in QW \cite%
{Hawrylak1995,Varga1999,RPV2000,Stebe20001,Varga2001,Dacal2002,Lozovik2004,BonitzCD}%
, quantum dots (QD) \cite%
{OgawaTakagahara,Klimov1994,Patton,Szafran2005,Peng2010,Kaniber,Jovanov},
quantum nanotubes \ \cite{Pedersen,Pedersen2,
Kammerlander,Pedersen3,DifMC,Matsunaga,Santos,Bondarev,Watanabe,Pedersen4,ColombierPRL,Yuma,Bondarev2,Bondarev3}
and quantum wires \cite{QW1 Baars, Zimmermann, Tsu2001, 1DEXP, ExQW,
PeetersPRB05, PeetersFewBody2006, PeetersPRB2008, Piermarocchi, QW 2012 Phys
Lett}. We cited these articles, but the recent literature on the subject is
not limited by them. The reduced dimensionality considerably increases the
binding energy of trions and biexcitons and, thus facilitates the formation
of these exciton complexes in semiconductor quantum wells, quantum wires
with different confinement geometry, quantum nanotubes and quantum dots.

Whilst in bulk materials excitonic characteristics are defined by dielectric
constant and effective masses of electrons and holes, in quantum
nanostructures, new controlling parameters, such as size, shape and material
distribution profile become crucial. According to numerous investigations
the main trend that has been revealed is that with size reduction binding
energy of excitons and excitonic complexes, as well as their decay
probability are strongly enhanced \cite{Chia,Nanoscale2014,Pokutnii}, which
seems very promising, and is crucial for applications in optoelectronics 
\cite{Wang2012}, such as light-emitting diodes \cite{Cheng2014},
photovoltaics \cite{Furchi2014}, and phototransistors \cite{Lembke2013}, to
cite just a few.


However, the situation is not always so straightforward; e.g. in Ref. \cite%
{Sarkar} it was reported that with decreasing size biexciton binding energy
was dropped to zero in InAs/AlAs quantum dots. Much more challenging is the
fact, that, with size reduction, some nonradiative processes also become
more intensive. In nanocrystal quantum dots/rods, nonradiative carrier
losses are dominated by surface trapping and multiparticle Auger relaxation 
\cite{Mikhailovsky}. Nonradiative processes connected to surface traps
increase with the size reduction due to the increase in surface-to-volume
ratio. However, these processes can be suppressed by using core-shell
structures for passivating surface traps. Auger recombination is a
nonradiative process in which the electron--hole recombination energy is
transferred to a third particle \cite{Kurzmann}. Auger recombination has a
relatively low efficiency in bulk semiconductors. However, Auger decay is
greatly enhanced in quantum-confined systems, which is mainly connected to
the increase of particle wave functions overlap and breakdown of
translational symmetry \cite{Grim}. Auger recombination is effective at high
excitation regime when number of excited carrier pairs exceeds one per
quantum dot \cite{Nozik} and when multi-exciton effects are intensive as
well. This is why trions and biexcitons still are subjects of extensive
investigations \cite{KezFBS2017,Falko,Falko2,KezerashviliTrion2018,Kez2018}.

Recent studies of one-dimensional (1D) nanostructures as nanotubes \cite%
{Santos,Yuma,Bondarev2,Bondarev3} and nanowires \cite{PeetersPRB2008,Efros
2010,QW 2012 Phys Lett,Xiong 2013} \ show that the trion and biexciton
binding energy depend on the electron to hole mass ratio and the geometric
characteristics of a nanostructure. Although the exciton complexes like
trions in solid state physics are very similar to the few-body bound systems
in atomic and nuclear physics there is a major difference related to band
effects, which make the effective masses of the electrons and holes smaller
than the bare electron mass, and screening effects, resulting from the host
lattice, which make the Coulomb force much weaker than in atomic systems.

The majority of recent research is conducted with core/shell NWs in which
the emitting core is overcoated with a thin layer of a semiconductor
material that has much higher band gap which enhances carrier localization
and suppresses Auger recombination. In group II--IV materials ZnO/ZnMgO,
CdSe/ZnS and CdSe/CdS are prime examples of such core/shell NWs which are
the subject of the present study.

In this paper, we present a theoretical approach to study a trion and a
biexciton in a NW in the framework of the effective-mass model using
Born-Oppenheimer approximation. We consider the formation of trions and
biexciton under the action of both the lateral confinement and the
localization potential. \emph{\ }Our approach allows us to obtain analytical
expressions for the binding energy and eigenfunctions of the trion and
biexciton. The corresponding energies are expressed by means of matrix
elements of effective one-dimensional cusp-type Coulomb potentials whose
parameters are determined self-consistently by employing the same
eigenfunctions of the confined electron and hole states. We calculated the
exciton, trion and biexciton binding energies in ZnO/ZnMgO, CdSe/ZnS and
CdSe/CdS core/shell NWs of cylindrical shape and their dependence on NW
radius. Because of high exciton binding energy ZnO/ZnMgO, CdSe/ZnS and
CdSe/CdS are a very good candidates for achieving efficient excitonic laser
action at room temperature and Auger recombination is expected to be reduced
for this class of materials \cite{Nitrid Semiconductor} as well as for
elongated core-shell structures \cite{Klimov3}. Having chosen the system
where Auger recombination is expected to be low, we aimed to find size
dependence of trion and biexciton binding energies in order to define
size/composition optimal for effective lasing. We also investigated the
process of autoionization during which the biexciton transforms to a trion.
We propose that as numbers of photo generated electrons and holes are the
same, this process should be the main source of generation of trions, which
in its turn are suspected to be the main reason of photoluminescence
intermediacy and efficiency drop \cite{Alicia2009}.

The paper is organized in the following way. In Sec.~\ref{Theory}, we
provide the theoretical model for a trion and biexciton in a core-shell NW
and obtain single-particle wavefunctions for confined electrons and holes,
which allows us to find the effective electron$-$hole, hole$-$hole and
electron$-$electron interactions in 1D. Using these interactions in the
framework of the effective-mass model we solve the one-dimensional Schr\"{o}%
dinger equations within the fixed center approximation for an exciton, trion
and biexciton and obtain the analytical expressions for the binding energies
and wavefunctions for these exciton complexes in Subsections \ref{Exc}, \ref%
{Tri}, and \ref{BiExc}, respectively. In Sec. \ref{Results} the results of
calculations and discussion are presented. Conclusions follow in Sec.~\ref%
{Conclusions}.

\section{Theoretical formalism}

\label{Theory}

\subsection{Setting the model}

\label{Model} We consider a formation of trions and biexcitons in a
core-shell nanowire. The system represents a cylindrical core of radius $a$,
surrounded by a cylindrical \ shell of thickness $b.$ The trion and
biexciton are a three- and four-body system and the corresponding Schr\"{o}%
dinger equations cannot be solved analytically, while a solution of the
Faddeev equations for a few-body system widely used in nuclear and atomic
physics is a challenging task, which involves complex numerical
computations. To overcome this difficulty, for the core-shell NW we consider
the theoretical model that is based on two assumptions:

\textbullet \qquad Coulomb interaction is assumed to be decisive only along
the NW axis and in radial direction motion of carriers is governed by the
strong lateral confinement perpendicular to the NW.

\textbullet \qquad Heavy holes on the average, move appreciably more slowly
than the electrons, which allows one to use the\ Born-Oppenheimer
approximation and solve Schr\"{o}dinger equation for fixed interhole
distances.

To solve the problem of a positive trion (two holes and one electron) and
biexciton (two holes and two electrons) laterally confined in a quantum NW
we adopt the Born-Oppenheimer approximation, very well known in physics of
molecules \cite{Atkins Friedman}. The Born-Oppenheimer approximation \cite%
{BornOppen} accounts for a difference in masses of light and heavy particles
and assumes that the light particles can respond almost instantly to heavy
particles' displacement \cite{Heitler and London}. The best example for a
such system is a hydrogen molecular ion H$_{2}^{+}$ and a hydrogen molecule H%
$_{2}$, which as a positively charged trion and biexciton consist from two
heavy and one light and two heavy and two light particles, respectively \cite%
{Atkins Friedman,Davidov}. Therefore, instead of solving the three-body Schr%
\"{o}dinger equation\ for all particle simultaneously one can treat heavy
particles as motionless and solve the Schr\"{o}dinger equation for a
definite position of heavy particles, taking the interparticle separation as
a parameter $R$. After that calculations are carried out for different $R$.

The application of the Born-Oppenheimer approximation naturally separates
the calculation into the following steps: due to the strong lateral
confinement perpendicular to the NW one first calculates the two-dimensional
(2D) energies and wave functions of the electron and hole, while neglecting
the Coulomb interaction between them. Therefore, the fast transverse motions
of charge carriers remain independent of each other. Next, using these wave
functions of transverse electron and hole motion, one can average the
three-dimensional (3D) Coulomb potential to a 1D Coulomb interaction between
the charge carriers along the NW. Finally, after an appropriate modeling of
these potentials by functions which depend on the distance between the
charge carriers, one should find energies and wavefuctions for a trion or a
biexciton for each fixed position of the holes by solving the corresponding
reduced 1D Schr\"{o}dinger equations.\emph{\ }

\begin{figure}[h]
\includegraphics[width=14.0cm]{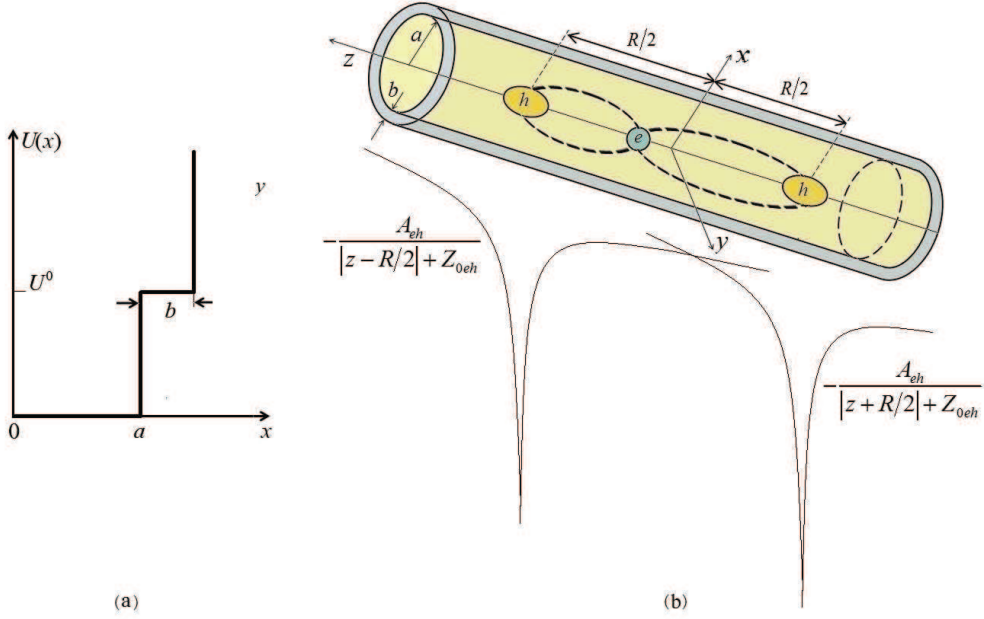} 
\caption{(Color online) (a) A lateral confinement potential. (b) Schematic
of the two 1D excitons sharing the same electron in a nanorwire to form a
positive trion. Schematically the electron in the field of two
one-dimensional cusp-type Coulomb potentials of the holes is shown in Fig.
1b.}
\label{Fig1}
\end{figure}

\label{excit}Let us assume that the conduction and highest valence bands are
decoupled, which is a reasonable approximation for the below considered type
of a NW because of the large direct band gap of the ZnO/ZnMgO, CdSe/ZnS and
CdSe/CdS materials. The full Hamiltonian for a three- or four-particle
excitonic complexes in a confinement within the single-band effective-mass
approximation can be written as

\begin{equation}
H=-\frac{\hbar ^{2}}{2}\sum\limits_{i=1}^{N}\frac{1}{m_{i}}\nabla
_{i}^{2}+V+\sum\limits_{i=1}^{N}U(\mathbf{r}_{i}),  \label{Hamiltonian}
\end{equation}%
where $r_{i}$ and $m_{i}$ are the position and effective mass of the $i$-th
particle, correspondingly, $N$ is a number of particle that is equal to 3
and 4 for the trion and biexciton, respectively, and $U(\mathbf{r}_{i})$ are
confinement potentials for each of particles. For example, one can consider
a lateral confinement potential shown in Fig. ~\ref{Fig1}a for a core-shell
NW. In the Hamiltonian (\ref{Hamiltonian}) $V=\sum\limits_{i<j}^{N}V(\mathbf{%
r}_{i},\mathbf{r}_{j})$ describes the Coulomb interactions between particles
with the electric charge $e$ in a medium with a dielectric constant $%
\varepsilon $ and is

\begin{equation}
V=\frac{e^{2}}{\varepsilon \left\vert \mathbf{r}_{1}-\mathbf{r}%
_{2}\right\vert }-\frac{e^{2}}{\varepsilon \left\vert \mathbf{r}_{1}-\mathbf{%
r}_{3}\right\vert }-\frac{e^{2}}{\varepsilon \left\vert \mathbf{r}_{2}-%
\mathbf{r}_{3}\right\vert }  \label{CoulombTrion}
\end{equation}%
in case of the positively charged trion and

\begin{equation}
V=\frac{e^{2}}{\varepsilon \left\vert \mathbf{r}_{1}-\mathbf{r}%
_{2}\right\vert }+\frac{e^{2}}{\varepsilon \left\vert \mathbf{r}_{3}-\mathbf{%
r}_{4}\right\vert }-\frac{e^{2}}{\varepsilon \left\vert \mathbf{r}_{1}-%
\mathbf{r}_{3}\right\vert }-\frac{e^{2}}{\varepsilon \left\vert \mathbf{r}%
_{2}-\mathbf{r}_{3}\right\vert }-\frac{e^{2}}{\varepsilon \left\vert \mathbf{%
r}_{2}-\mathbf{r}_{3}\right\vert }-\frac{e^{2}}{\varepsilon \left\vert 
\mathbf{r}_{2}-\mathbf{r}_{4}\right\vert }
\end{equation}%
in case of the biexciton.

Below we consider the formation of trions and biexcitons in a core-shell NW.
The system represents a cylindrical core of radius $a$, surrounded by a
shell of thickness $b$, as schematically shown in Fig.~\ref{Fig1}b. To find
the binding energies, calculate the energy spectra of trions and biexcitons
and find eigenfunctions needed for calculation their optical properties, we
must solve the Schr\"{o}dinger equation with Hamiltonian (\ref{Hamiltonian}%
). We assume that the lateral confinement is strong, so that only the lowest
subband for the electron and hole is occupied. This assumption allows a
reduction of the Schr\"{o}dinger equation to an effective one-dimensional
form. Assuming strong lateral confinement, we are allowed to separate the $z$
motion from the lateral motion in the $xy$ plane. In other words, we assume
that the Coulomb interaction does not affect the $xy$ motion of the
particles, so that we can separate the electron and hole motion confined in
the lateral direction from the electron-hole relative motion. Therefore, the
envelope function for a trion or biexciton can be approximated as

\begin{equation}
\Psi (\mathbf{r}_{1},\mathbf{r}_{2},...,\mathbf{r}_{N})=\Phi
(z_{1},z_{2},...,z_{N})\prod\limits_{i}\psi _{i}(\rho _{i},\varphi _{i}),
\label{Trial Function}
\end{equation}%
where $\Phi (z_{1},z_{2},...,z_{N})$ is the envelope function describing the
electron-hole relative motion in the trion $(N=3)$ or biexciton $(N=4)$ in
the $z$ axis along the NW and $\psi _{e}(\rho _{e},\varphi _{e})$ $(\psi
_{h}(\rho _{h},\varphi _{h}))$ is the radial single-particle wave functions
for an electron (hole) for the lateral motion in the $xy$ plane, which due
to the axial symmetry of the system depend on cylindrical coordinates for
each of particles. Here and below the Cartesian coordinates, wavefunctions
and masses of the electron (hole) are denoted with the suffix $e$ and $h$,
respectively.

By averaging the Schr\"{o}dinger equation with Hamiltonian (\ref{Hamiltonian}%
) by using function (\ref{Trial Function}) after the separation of variables
we obtain 
\begin{equation}
-\frac{\hbar ^{2}}{2m_{e(h)}}\left[ \frac{1}{\varrho }\frac{\partial }{%
\partial \varrho }\left( \varrho \frac{\partial }{\partial \varrho }\right) +%
\frac{1}{\varrho ^{2}}\frac{\partial ^{2}}{\partial \varphi ^{2}}\right]
\psi _{e(h)}(\rho ,\varphi )+U_{e(h)}(\varrho )\psi _{e(h)}(\rho ,\varphi
)=E_{e(h)}\psi _{e(h)}(\rho ,\varphi ),  \label{Single particle}
\end{equation}%
which is the Schr\"{o}dinger equation of single-particle states for
electrons and holes confined in a NW and the following Hamiltonian

\begin{equation}
H_{X^{+}}=-\frac{\hbar ^{2}}{2m_{h}}\left[ \frac{\partial ^{2}}{\partial
z_{1h}^{2}}+\frac{\partial ^{2}}{\partial z_{2h}^{2}}\right] -\frac{\hbar
^{2}}{2m_{e}}\frac{\partial ^{2}}{\partial z_{3e}^{2}}%
+V_{eh}^{eff}(z_{1h}-z_{3e})+V_{eh}^{eff}(z_{2h}-z_{3e})+V_{hh}^{eff}(z_{1h}-z_{2h}),
\label{Trion}
\end{equation}%
in the case of trion, while in the case of biexciton the corresponding
Hamiltonian reads 
\begin{eqnarray}
H_{XX} &=&-\frac{\hbar ^{2}}{2m_{h}}\left[ \frac{\partial ^{2}}{\partial
z_{1h}^{2}}+\frac{\partial ^{2}}{\partial z_{2h}^{2}}\right] -\frac{\hbar
^{2}}{2m_{e}}\left[ \frac{\partial ^{2}}{\partial z_{3e}^{2}}+\frac{\partial
^{2}}{\partial z_{4e}^{2}}\right]
+V_{eh}^{eff}(z_{1h}-z_{3e})+V_{eh}^{eff}(z_{1h}-z_{4e})  \nonumber \\
&&+V_{eh}^{eff}(z_{2h}-z_{3e})+V_{eh}^{eff}(z_{2h}-z_{4e})+V_{ee}^{eff}(z_{3e}-z_{4e})+V_{hh}^{eff}(z_{1h}-z_{2h}).
\label{Biexciton}
\end{eqnarray}

In Hamiltonians (\ref{Trion}) and (\ref{Biexciton}) $V_{eh}^{eff}$, $%
V_{hh}^{eff}$ and $V_{ee}^{eff}$ are effective electron$-$hole, hole$-$hole
and electron$-$electron interactions that are defined in Appendix \ref{app:A}%
. Conceptually a positively charged 1D trion can be considered as two
ground-state 1D excitons sharing the same electron to form a positive trion
state and 1D biexciton as two 1D excitons that are sharing the same two
interacting electrons to form a biexciton bound state. A schematic of this
concept for the trion is depicted in Fig.~\ref{Fig1}b.

Thus, finding the eigenfunctions and eigenenergies for the trion and
biexciton for the Hamiltonian (\ref{Hamiltonian}) is reduced to the solution
of the Schr\"{o}dinger equation for single-particle states of the electrons
and the hole confined in a NW (\ref{Single particle}) and a solution of the
Schr\"{o}dinger equation with the Hamiltonian (\ref{Trion}) for the trion
and Hamiltonian (\ref{Biexciton}) for the biexciton with the effective
potentials (\ref{Veh}) - (\ref{Vee}). The effective potentials (\ref{Veh}) -
(\ref{Vee}) between charged particles are free from the singularity of the
bare Coulomb potential at the origin as a result of averaging with the
lateral subband wave functions. However, these effective potentials can be
given only numerically. Following Ogawa and Takagahara \cite{OgawaTakagahara}
the effective potentials (\ref{Veh}) - (\ref{Vee}) for 1D semiconductors
usually are modeled by effective one-dimensional cusp-type Coulomb
potentials approximated by the first order rational function $\frac{A}{%
(z+Z_{0})}$, where $z$ is interparticle distance in $z$-direction and $A$
and $Z_{0}$ are fitting parameters. These parameters are defined by wave
functions $\psi _{e(h)}(\rho ,\varphi )$ that are the solutions of (\ref%
{Single particle}), which in its turn, depend on a NW geometry, particles'
effective masses, and band offsets $U_{e(h)}$. Consequently, they vary with
the NW radius and are different for the electron$-$hole, hole$-$hole and
electron$-$electron interactions (see Appendix \ref{app:A}).

\subsection{Single-particle states for confined electrons and holes}

\label{Single E and H} Let us solve the Schr\"{o}dinger equation (\ref%
{Single particle}) for single-particle states of the electron and the hole
confined in a core-shell NW. We assume that confinement potentials for the
electron and hole depicted in Fig.~\ref{Fig1}a have a stepped well shape,
are different and have an axial symmetry

\begin{equation}
U(\rho )=\left\{ 
\begin{array}{c}
0,\text{ when }\rho <a, \\ 
U_{e(h)}^{0},\text{ when }a\leq \rho \leq a+b, \\ 
\infty ,\text{ \ when }\rho >a+b,%
\end{array}%
\right.  \label{ConfinPotential}
\end{equation}%
where $U_{e(h)}^{0}$ is the conduction (valence) band offset between core
and shell materials and, obviously, it is different for electrons and holes
due to the different bands that leads to their different masses. Proceeding
from the cylindrical symmetry the solution of Eq. (\ref{Single particle})
should have the form 
\begin{equation}
\psi _{e(h)}(\rho ,\varphi )=\phi _{e(h)}(\rho )e^{in\varphi },\text{ }%
n=0,\pm 1,\pm 2,...\text{ .}  \label{CylinderSolution}
\end{equation}%
By substituting (\ref{CylinderSolution}) into (\ref{Single particle}) we
obtain 
\begin{equation}
-\frac{\hbar ^{2}}{2m_{e(h)}}\left[ \frac{1}{\rho }\frac{\partial }{\partial
\rho }\left( \rho \frac{\partial }{\partial \rho }\right) +\frac{m^{2}}{\rho
^{2}}\right] \phi _{e(h)}(\rho )+U(\varrho )\phi _{e(h)}(\rho )=E_{e(h)}\phi
_{e(h)}(\rho ).
\end{equation}%
The solutions of this equation are Bessel functions: within the core we have
the Bessel function of first kind $J_{n}$, which is finite at $\rho =0$,
while within the shell the solution is a linear combination of the Bessel
functions of second type $K_{n}$ and $I_{n}$

\begin{equation}
\phi _{e(h)}(\rho )=\left\{ 
\begin{array}{c}
CJ_{n}(k\rho ),\text{ when }\rho <a, \\ 
C_{1}K_{n}(\varkappa \rho )+C_{2}I_{n}(\varkappa \rho ),\text{ when }a\leq
\rho \leq a+b,%
\end{array}%
\right.  \label{RadialSolution}
\end{equation}%
where 
\begin{equation}
k^{2}=\frac{2m_{e(h)}E_{e(h)}}{\hbar ^{2}},\text{ \ \ \ \ }\varkappa ^{2}=%
\frac{2m_{e(h)}U_{e(h)}^{0}}{\hbar ^{2}}-k^{2}.
\end{equation}%
The $C,$ $C_{1}$ and $C_{2}$ coefficients in (\ref{RadialSolution}) are
defined from the condition of continuity and smoothness of the wavefunction
at the boundary, and from the condition of its orthonormality. Consequently,
the energy of radial motion of electrons (holes) are defined by means of
equation $C_{1}K_{n}(\varkappa b)+C_{2}I_{n}(\varkappa b)=0.$ Having found $%
k $, we can define the in-plane $E_{e(h)}$ energy and the wave functions of
a confined non-interacting electron and hole, respectively. The solutions $%
\psi _{e(h)}(\rho ,\varphi )$ of a single-particle radial Schr\"{o}dinger
equation (\ref{Single particle}) are used to average three-dimensional
potentials in (\ref{Veh}) - (\ref{Vee}) over in-plane motion. As a result,
one-dimensional effective potentials are obtained for the electron$-$hole,
hole$-$hole and electron$-$electron interactions that then are parameterized
in the form $\frac{A}{(z+Z_{0})}$ \ as shown in Appendix \ref{app:A}.

\subsection{Exciton in a nanowire}

\label{Exc} In the ideal limit the 1D electron-hole system with a perfect
confinement, can be treated as a "one-dimensional hydrogen-atom" problem in
the framework of the effective-mass approximation. As a first step let us
consider the interacting electron and hole in a 1D NW and model an exciton
by using an effective one-dimensional cusp-type Coulomb potential. Following 
\cite{OgawaTakagahara} one can write the equation that describes relative
motion of the electron and hole bound with the cusp-type Coulomb interaction
in one-dimensional radially confined NW in the form

\begin{equation}
-\frac{\hbar ^{2}}{2\mu }\frac{d^{2}\Phi _{X}(z)}{dz^{2}}-\frac{A_{eh}}{%
\left\vert z\right\vert +Z_{0eh}}\Phi _{X}(z)=E_{X}\Phi _{X}(z).
\label{Exciton}
\end{equation}%
Here $\mu $ is the reduced effective mass of electron-hole pair, $A_{eh},$ $%
Z_{0eh}$ are the fitting parameters for the effective electron$-$hole
one-dimensional cusp-type Coulomb potentials obtained through the
parametrization of Eqs. (\ref{Veh}), $z=z_{e}-z_{h}$ is the relative
electron-hole motion coordinate, $E_{X}$ is the binding energy of the
electron and hole that form the exciton and $\Phi _{X}(z)$ is the
corresponding wavefunction. Eq. (\ref{Exciton}) has the same form as the
equation for one-dimensional hydrogen atom studied by Loudon \cite{Loudon1}.
One can introduce the following notations

\begin{equation}
\xi ^{2}=\frac{\hbar ^{2}\eta _{0}^{2}}{2\mu E_{x}},\text{ }\eta _{0}=\frac{%
A\mu }{\hbar ^{2}},\text{ }x=\frac{2\eta _{0}(\left\vert z\right\vert
+Z_{eh})}{\xi }  \label{Notation}
\end{equation}%
and reduce (\ref{Exciton}) to the Whittaker's equation

\begin{equation}
\frac{d^{2}\zeta (x)}{dx^{2}}+\left[ -\frac{1}{4}+\frac{\zeta (x)}{x}\right]
=0.  \label{Whittaker}
\end{equation}%
The solution of (\ref{Whittaker}) is the Whittaker function $W_{\xi
,1/4}(x), $ $\zeta (x)=W_{\xi ,1/4}(x),$ as shown in Refs. \cite{Loudon1,
Loudon2}. The value of $\xi $ which defines $E_{X}$\ and $\Phi _{X}(z)$, is
determined by the boundary condition stating that for even states the
derivative of wave function at $z=0$ must turn to zero

\begin{equation}
\frac{d}{dz}\left[ W_{\xi ,1/4}\left( \frac{2\eta _{0}(\left\vert
z\right\vert +Z_{eh})}{\xi }\right) \right] _{z=0}=0.  \label{ExcitEner}
\end{equation}%
The energy of exciton in a quantum NW is a sum of energies of radial motion
of electron and hole $E_{e}+E_{h}$, and the energy of the exciton $E_{X}.$


\subsection{Positive trion in a nanowire}

\label{Tri} Following \cite{Atkins Friedman} for positively charged trions
bound by the effective one-dimensional cusp-type Coulomb potentials the
Hamiltonian (\ref{Trion}) in the Born-Oppenheimer approximation can be
written as

\begin{equation}
H_{X^{+}}=-\frac{\hbar ^{2}}{2\mu }\frac{d^{2}}{dz^{2}}-\frac{A_{eh}}{%
\left\vert z-R/2\right\vert +Z_{0eh}}-\frac{A_{eh}}{\left\vert
z+R/2\right\vert +Z_{0eh}}+\frac{A_{hh}}{R+Z_{0hh}},
\label{Trion_Born-Oppenheimer}
\end{equation}%
where $\mu $ is electron-hole reduced mass, $R$\ is distance between two
holes, which are assumed to be motionless at $z=\pm R/2$, and $A_{eh},$ $%
Z_{0eh}$ and $A_{hh},$ $Z_{0hh}$ are the fitting parameters for the
effective electron$-$hole and hole$-$hole one-dimensional cusp-type Coulomb
potentials obtained through the parametrization of Eqs. (\ref{Veh}) and (\ref%
{Vhh}), correspondingly. Schematically the electron in the field of two 1D
cusp-type Coulomb potentials is shown in Fig.~\ref{Fig1}b.

For solution of the Schr\"{o}dinger equation for the trion with Hamiltonian (%
\ref{Trion_Born-Oppenheimer}) we use the method of linear combination of
atomic orbitals (LCAO) \cite{Atkins Friedman}. In the framework of \ the
LCAO method the eigenfunction of trion is presented as a linear combination
of single exciton wave functions centered at $z=\pm R/2$

\begin{equation}
\Phi =c_{1}\Phi _{X_{1}}+c_{2}\Phi _{X_{2}},  \label{Linear}
\end{equation}%
where $\Phi _{X_{1}}$ and $\Phi _{X_{2}}$ are the solutions of the Schr\"{o}%
dinger equation

\begin{equation}
-\frac{\hbar ^{2}}{2\mu }\frac{d^{2}\Phi _{X_{1}(X_{2})}(z)}{dz^{2}}-\frac{%
A_{eh}}{\left\vert z\pm R/2\right\vert +Z_{0eh}}\Phi
_{X_{1}(X_{2})}=E_{X}\Phi _{X_{1}(X_{2})}.
\end{equation}%
Close to the hole located at $R/2$ the wavefunction $\Phi _{X_{1}}$ will
resemble a single electron orbital, while the wavefunction $\Phi _{X_{2}}$
will represent the electron orbital near the hole located at $-R/2$. Thus,
the linear combination (\ref{Linear}) represents both cases. The states $%
\Phi _{X_{1}}$ and $\Phi _{X_{2}}$ are degenerate due to symmetry. Following
the well-known perturbation theory for the degenerate states \cite{Landau},
one can obtain the normalized wavefunctions for the first two states 
\begin{equation}
\Phi _{\pm }=\frac{1}{\sqrt{2(1\pm S)}}(\Phi _{X_{1}}\pm \Phi _{X_{2}}),
\end{equation}%
where $S=\left\langle \Phi _{X_{1}}\right\vert \left. \Phi
_{X_{2}}\right\rangle $ is an overlap integral. The wavefunction $\Phi _{+}$
corresponds to the localization of the electron density between holes
(bonding orbital) and the accumulation of electron density in the interhole
region is simulated due to the constructive interference that takes place
between the two electron waves centred on neighboring holes, while the
wavefunction $\Phi _{-}$ \ has a node between the holes (antibonding
orbital). Trion energies corresponding to these states are:

\begin{eqnarray}
E_{X^{+}}^{+} &=&E_{X}+\frac{J+K}{1+S}+E_{hh},  \label{TrionEnergy1} \\
E_{X^{+}}^{-} &=&E_{X}+\frac{J-K}{1-S}+E_{hh},  \label{TrionEnergy2}
\end{eqnarray}%
where the expressions for $J$ and $K$ are given in Appendix \ref{app:B}. The
last term in (\ref{TrionEnergy1}) and (\ref{TrionEnergy2}) is the energy of
interaction between two holes

\begin{equation}
E_{hh}=\frac{A_{hh}}{R+Z_{0hh}}.  \label{HHenergy}
\end{equation}%
Thus, to find the dependence of binding energy of the trion on interhole
separation $R$\ one needs to evaluate the relevant matrix elements $J$ and $%
K $ and calculate the interaction energy between two holes (\ref{HHenergy}).

\subsection{Biexciton in a nanowire}

\label{BiExc} Now let us consider a 1D biexciton for which the model of the
hydrogen molecule H$_{2}$ is used. Following \cite{Davidov} for the
biexciton, where two electrons and two holes are bound via the effective
one-dimensional cusp-type Coulomb potentials, the Hamiltonian (\ref%
{Biexciton}) in the Born-Oppenheimer approximation can be written as

\begin{eqnarray}
H_{XX} &=&-\frac{\hbar ^{2}}{2\mu }\left( \frac{d^{2}}{dz_{1}^{2}}+\frac{%
d^{2}}{dz_{2}^{2}}\right) -\frac{A_{eh}}{\left\vert z_{1}-R/2\right\vert
+Z_{0eh}}-\frac{A_{eh}}{\left\vert z_{2}-R/2\right\vert +Z_{0eh}}-\frac{%
A_{eh}}{\left\vert z_{1}+R/2\right\vert +Z_{0eh}}-\frac{A_{eh}}{\left\vert
z_{2}+R/2\right\vert +Z_{0eh}}+  \nonumber \\
&&\frac{A_{hh}}{R+Z_{0hh}}+\frac{A_{ee}}{z_{12}+Z_{0ee}},
\label{Biexciton Hamiltonian}
\end{eqnarray}%
where $z_{12}=\left\vert z_{1}-z_{2}\right\vert $ and $A_{ee}$ and $Z_{0ee}$
are the fitting parameters for the effective one-dimensional cusp-type
Coulomb potentials in the Hamiltonian (\ref{Biexciton}) for the electron$-$%
electron interaction obtained from (\ref{Vee}), while the other fitting
parameters are already defined in Subsec. \ref{Tri}. The wave function of
the biexciton can be constructed by means of single exciton wave functions
again. If the electron--electron and hole-hole interactions are ignored one
can use $\Phi _{X_{1}}(z_{1})\Phi _{X_{2}}(z_{2})$ as an approximation for
the biexciton wavefunction. However, the symmetry of the wavefunction has to
be taken into account. Following \cite{Davidov} we construct symmetric and
antisymmetric wave functions coinciding to antiparallel (singlet state) and
parallel orientation (triplet state) of spins of electrons as:

\begin{equation}
\Phi _{\pm }(z_{1},z_{2})=\frac{1}{\sqrt{2(1\pm S)}}\left[ (\Phi
_{X_{1}}(z_{1})\Phi _{X_{2}}(z_{2})\pm \Phi _{X_{1}}(z_{2})\Phi
_{X_{2}}(z_{1})\right] .
\end{equation}

The energies of bonding symmetric and antibonding antisymmetric states in
the first order perturbation theory are found as mean values of (\ref%
{Biexciton Hamiltonian}). The corresponding energies for these states are:

\begin{eqnarray}
E_{XX}^{+} &=&2E_{X}+\frac{Q+P}{1+S^{2}},  \label{Biexciton Energy1} \\
E_{XX}^{-} &=&2E_{X}+\frac{Q-P}{1-S^{2}},  \label{Biexciton Energy2}
\end{eqnarray}%
where $Q$ and $P$ are given in Appendix \ref{app:C}.

The binding energy of biexciton is defined as maximal value of $%
2E_{X}-E_{XX}(R)$, and when this difference is positive biexciton is a
stable system.

\section{\protect\bigskip Results of calculations and discussion}

\label{Results} Our calculations are based on the assumption that the
Coulomb interaction strength in the radial direction is much weaker than the
lateral confinement effect and the Coulomb interaction is decisive only
along the \textit{z}-axis and is presented by the 1D cusp-type Coulomb
potentials for the electron$-$hole, hole$-$hole and electron$-$electron
interactions. The Born-Oppenheimer approximation successfully used in the
physics of molecules assumes that the ratio of mass of the atomic nuclei and
electrons is large enough. In the case of positively charged trions and
biexcitons the ratio of hole-electron mass is much less than the
proton-electron mass ratio, which makes questionable not only the validity
of the Born-Oppenheimer approximation, but also the existence of bound
states. In \cite{Aven Prener} it is stated that exciton bound to
neutral/ionized donor for which models of H$_{2}^{+}$ and H$_{2}$ are also
used, the binding energy sharply increases when the hole-electron mass ratio
varies from 1 to 3. For II-VI semiconductors, where the mass ratio $%
m_{h}/m_{e}$ varies between 3$-$5, for calculation of trion and biexciton
states the Born-Oppenheimer approximation is often used \cite{AP}. Even in
the case when $m_{h}=m_{e}$ satisfactory values are obtained for the binding
energy of the two- and three-dimensional biexcitons \cite{Lozovik}.

We calculate the trion and biexciton binding energies in ZnO/ZnMgO, CdSe/ZnS
and CdSe/CdS core/shell quantum structures of a cylindrical shape. The
calculations are performed for the set of parameters listed in Table 1. Here
the effect of the differences in the effective masses in core and shell
material is not taken into account. For ZnO/ZnMgO and CdSe/CdS structures
this difference is not significant. In CdSe/ZnS structure the barrier height
is very high, penetration of carriers into the barrier is not substantial.
Because of high exciton binding energy ZnO/ZnMgO is a very good candidate
for achieving efficient excitonic laser action at room temperature.
ZnO/ZnMgO is a wide bandgap (3.4 eV) material, has sufficient hole-electron
mass ratio. The most important part of the ZnO/ZnMgO band structure, which
is the bottom of the conduction band and the top of the valence bands in the
vicinity of the $\Gamma -$point, can be well described within the
effective-mass approximation \cite{Zn Oxide}.

CdSe/ZnS and CdSe/CdS are also promising direct-bandgap II--VI
semiconductors active in the visible range, with potential applications in
electronics and optoelectronics \cite{Shalev2017}. Nanostructures on the
base of CdSe are one of the extensively investigated low dimensional
semiconductor structures with the aim of application field-effect
transistors, photodetectors, and light-emitting diodes \cite{Zhou2007}. The
bandgap of CdSe is lower than that of ZnO $-$ 1.74 eV \cite{Trindade1997}.
Exciton binding energy in bulk CdSe also is significantly lower than that in
bulk ZnO -- 0.13 eV \cite{Meulenberg2009}. However, investigations show
possibility of effective control/enhancement of stability of excitonic
complexes in nanostrudtures \cite{Meulenberg2009, Hawrylak2010}.

\begin{table}[b]
\caption{Input parameters for ZnO/ZnMgO and CdSe/ZnS NWs. $m_{e}$ and $m_{e}$
are the mass of the electron and hole, respectively, $U_{e}^{0}$ and $%
U_{h}^{0}$ are the lateral confinement potentials for a conduction and
valence band offset between core and shell materials, respectively, and $%
\protect\epsilon $ is dielectric constant. $m_{0}$ is the mass of free
electron. }
\label{t1}
\begin{center}
\begin{tabular}{cccccc}
\hline\hline
& $m_{e}/m_{0}$ & $m_{h}/m_{0}$ & $U_{e}^{0},$ eV & $U_{h}^{0},$ eV & $%
\varepsilon$ \\ \hline
ZnO/ZnMgO & 0.24 \cite{Kuzmina} & 0.86 \cite{Kuzmina} & 0.37 \cite{Janotti}
& 0.31 \cite{Janotti} & 8.13 \\ 
CdSe/ZnS & 0.13 \cite{Aven Prener} & 0.45 \cite{Aven Prener} & 1.2 \cite%
{Reiss} & 0.7 \cite{Reiss} & 10.2 \\ 
CdSe/CdS & 0.13 & 0.45 & 0.30 \cite{Reiss} & 0.44 \cite{Reiss} & 10.2 \\ 
\hline\hline
\end{tabular}%
\\[0pt]
\end{center}
\par
\end{table}

\begin{figure}[b]
\begin{center}
\includegraphics[width=8.75cm]{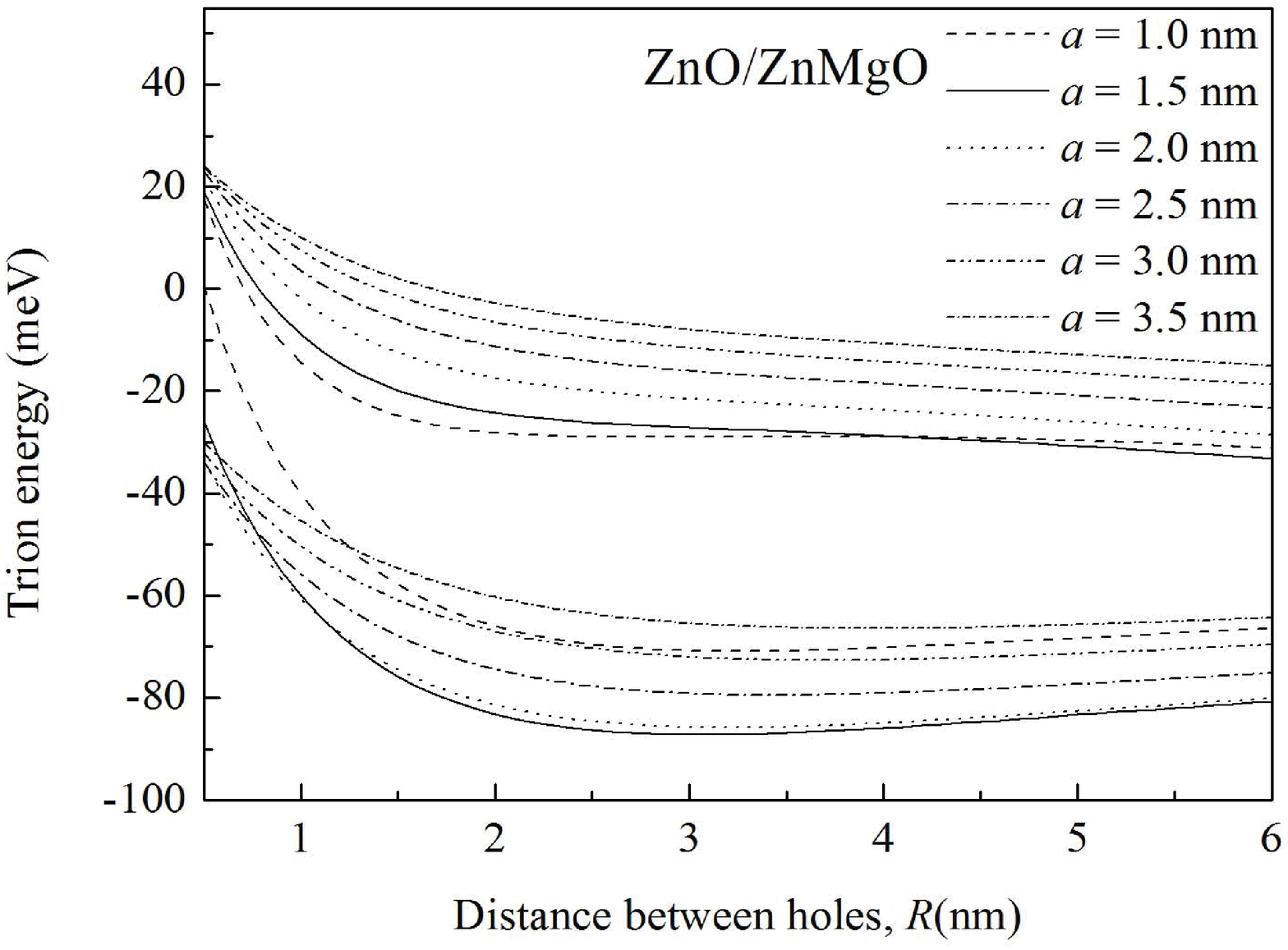} \vspace{-0.1cm} %
\includegraphics[width=8.75cm]{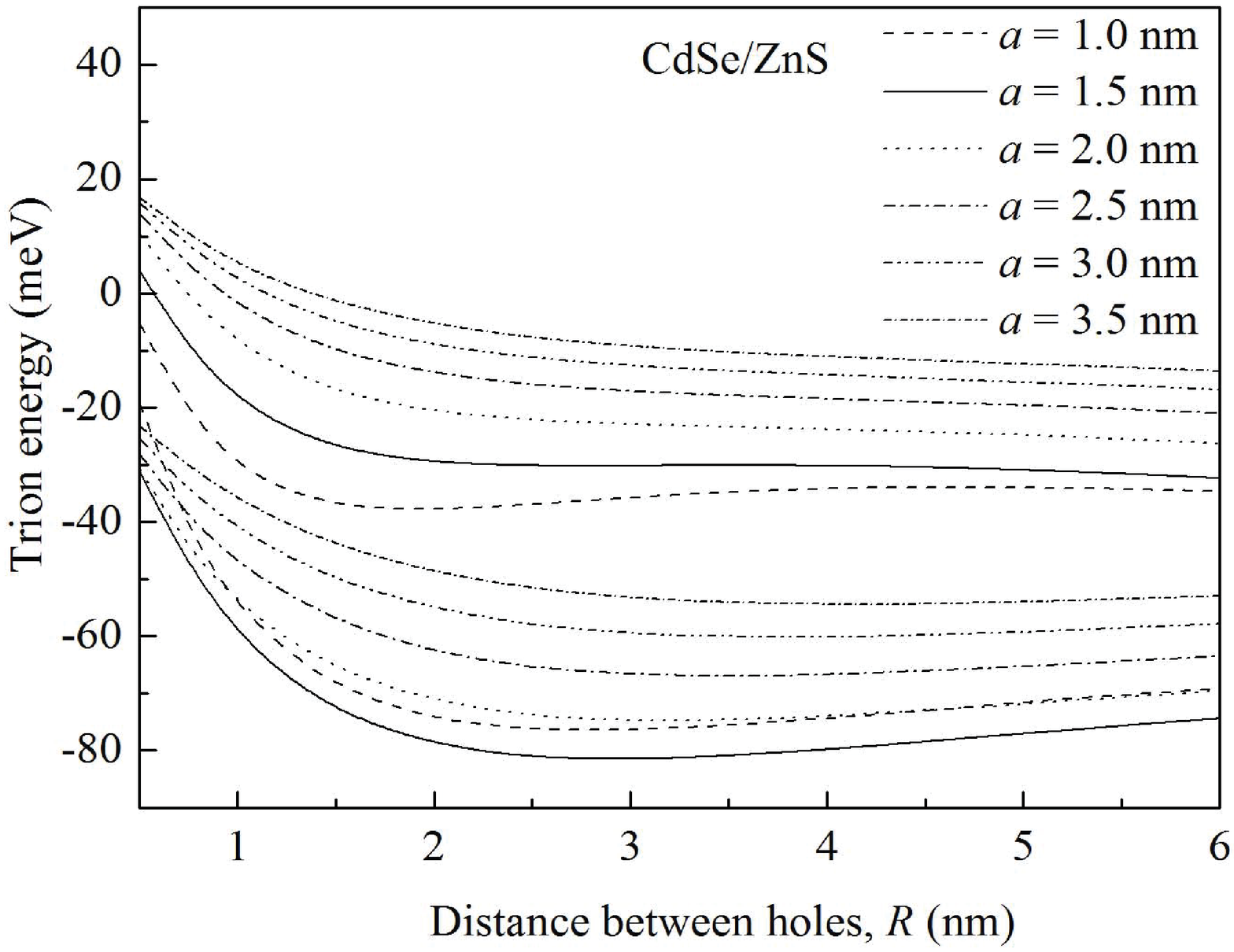} \vspace{-0.1cm}
\end{center}
\caption{ The dependence of trion energy in ZnO/ZnMgO and CdSe/ZnS NWs on
interhole distance for different radii of a NW. The thickness of the
surrounding dielectric shell is $b=2$ nm.}
\label{Fig3}
\end{figure}

\begin{figure}[b]
\begin{center}
\includegraphics[width=8.75cm]{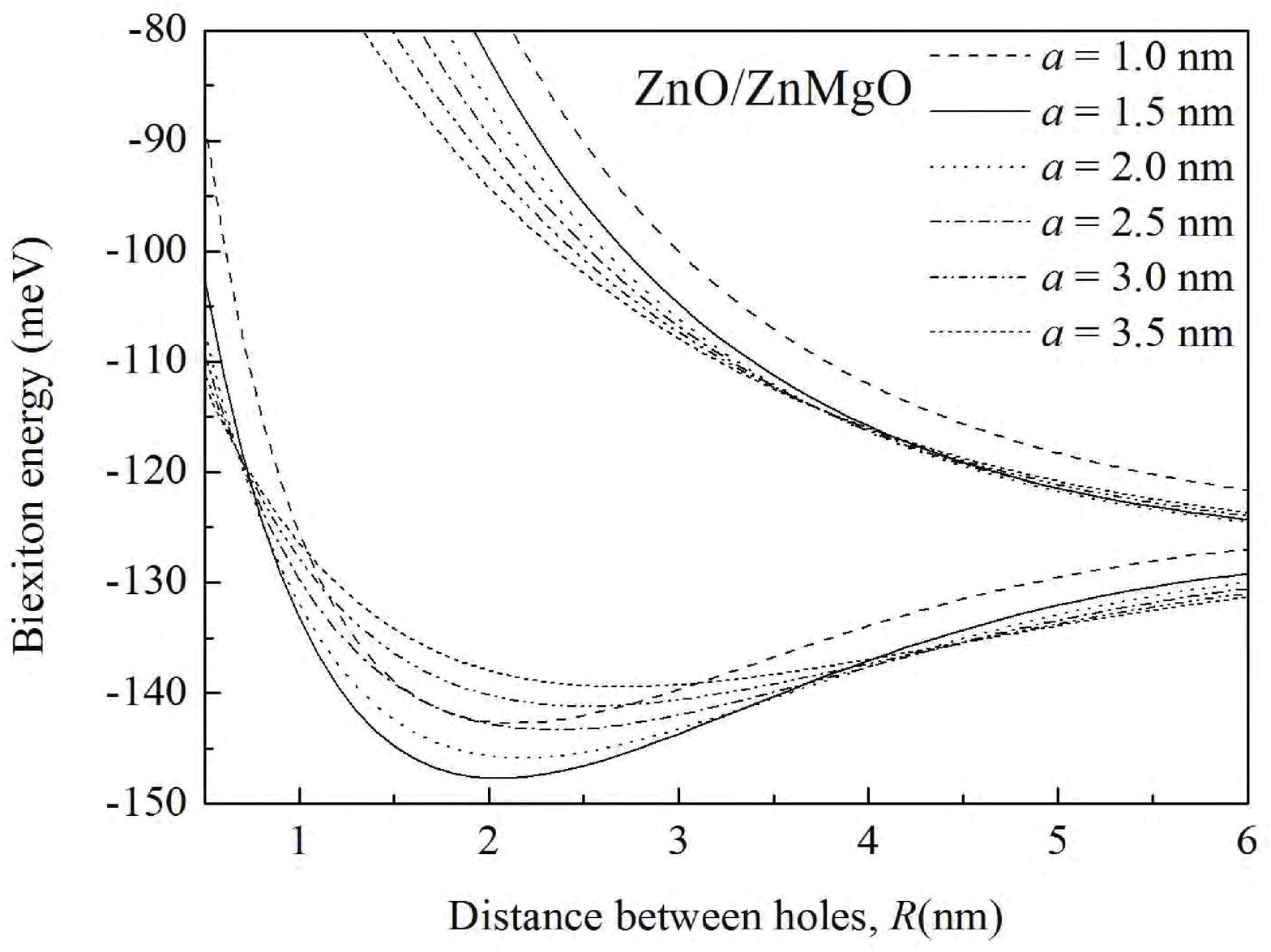} \vspace{-0.1cm} %
\includegraphics[width=8.75cm]{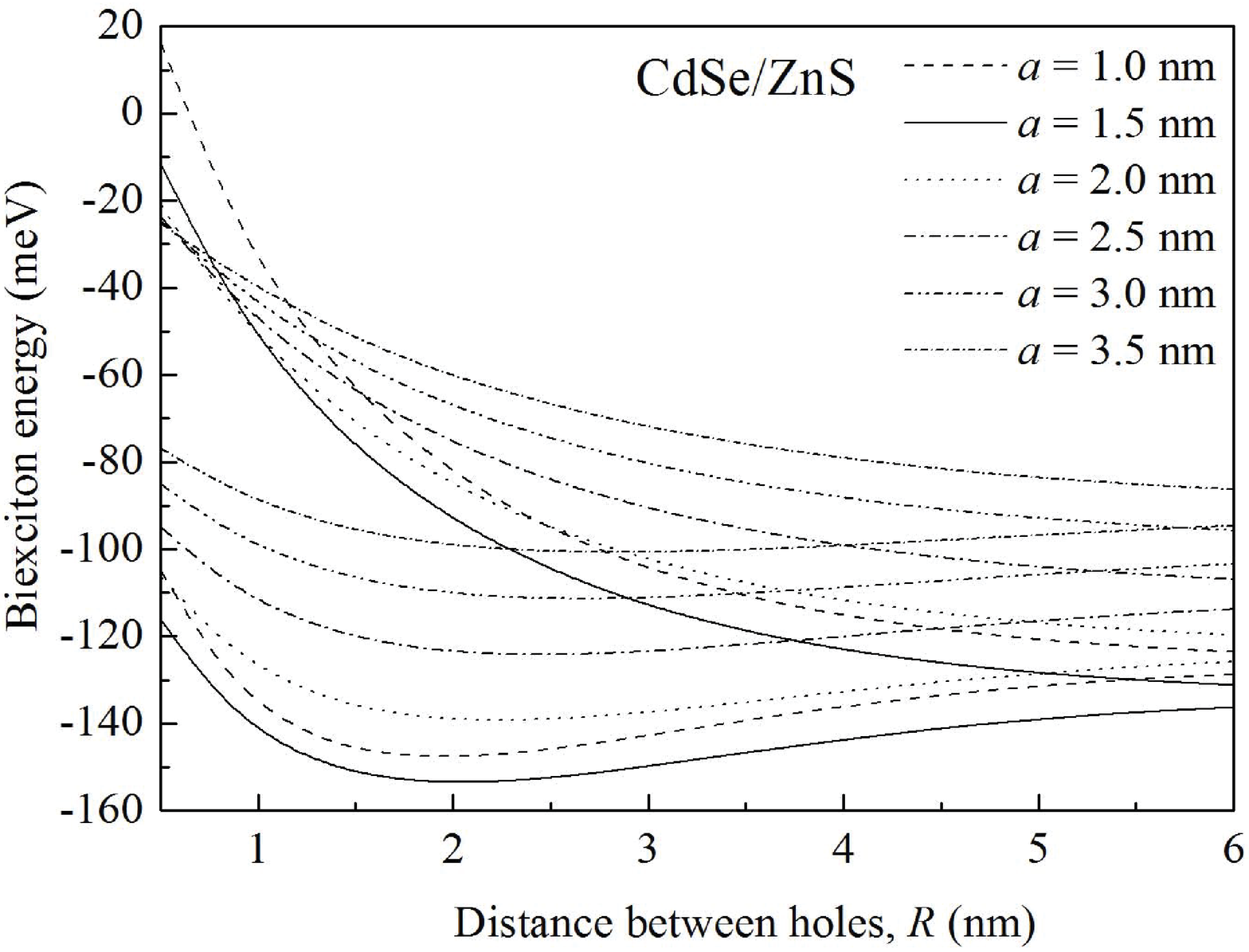} \vspace{-0.1cm}
\end{center}
\caption{ The dependence of biexciton energy in ZnO/ZnMgO and CdSe/ZnS NWs
on interhole distance for different radii of a NW. The thickness of the
surrounding dielectric shell is $b=2$ nm.}
\label{Fig5}
\end{figure}

As a first step we calculate the effective interactions (\ref{Veh}) - (\ref%
{Vee}) by averaging over the electron $\psi _{e}(\rho _{e},\varphi _{e})$
and hole $\psi _{h}(\rho _{h},\varphi _{h})$ wavefunctions, which reduce the
3D Coulomb potential to a one-dimensional potential that depends only on the
coordinate of two particle relative motion. The corresponding computational
modeling allows one to find the numerical values of the fitting parameters $%
A $ and $Z_{0}$ for the 1D cusp-type Coulomb potentials. Once these
constants are known one can use them as the input parameters and evaluate $J$%
\ and $K$\ given in (\ref{J}) and (\ref{K}), and using Eqs. (\ref%
{TrionEnergy1}) and (\ref{TrionEnergy1}) find the trion energy for given
interhole separation $R$. In the case of biexciton using the same matrix
elements $J$\ and $K$\ given in (\ref{J}) and (\ref{K}), and the matrix
elements $\mathfrak{J}$ and $\mathfrak{K}$ given in (\ref{j}) and (\ref{k}),
one can find $Q$ and $P$ defined in (\ref{Q P}) and determine the biexciton
energy by mean of Eqs. (\ref{Biexciton Energy1}) and (\ref{Biexciton Energy2}%
).

We study the dependence of the trion and biexciton binding energy on the
interhole distance, the NW radius and the thickness of the surrounding
dielectric shell using Eqs. (\ref{TrionEnergy1}) and (\ref{TrionEnergy2}),
and (\ref{Biexciton Energy1}) and (\ref{Biexciton Energy2}), respectively.
When the trion or biexciton energies in ZnO/ZnMgO, CdSe/ZnS and CdSe/CdS as
a function of interhole distance exhibit minimums at the particular
interhole distances and for the particular NW radius it can be a signature
of possible existence of the bound state of the trion or biexciton. If the
minimum lies below the energy of a separated exciton and a hole its
indicates the formation of the trion, while if the minimum energy is below
the energy of two separated excitons, it is an indication of the biexciton
formation.

The results of calculations for the trion energies in ZnO/ZnMgO and CdSe/ZnS
NWs as a function of interhole distance $R$\ are presented in Fig. \ref{Fig3}%
. The calculation is performed for different radii of a NW. The energy
curves vary with the interhole distance. The upper branches of curves
correspond to the antibonding orbital for the antisymmetric state, while the
lower branches of curves correspond to the bonding orbital for the symmetric
state. The steep rise dependence in energy at $R<1.5$\ for both ZnO/ZnMgO
and CdSe/ZnS NWs is largely due to the increase in the hole-hole potential
energy as the two holes are brought close together. When a curve exhibits
minimum and its energy is below the energy of a separated exciton and a hole
one can confirm the existence of a bound trion. For all values of a NW
radius the typical behavior of a symmetric bonding state for ZnO/ZnMgO NW is
observed with minimum at some particular hole-hole distance, while one can
see the minimum of the energy for the antisymmetric state only for the NW
with the smallest radius $a=1.0$ nm. The minimums are not sharp,
nevertheless bound states of trions exist. In the case of CdSe/ZnS the
curves for the trion energies as a function of interhole distance exhibit
minimum for the symmetric bonding states as well as for the antibonding
antisymmetric states. The minimums are pronounced for antisymmetric states
for the radius of NW $a=1.0$ nm and $a=1.5$ nm, however, there is no minimum
in the antibonding state for the other considered values of NW radii.

Let us now present results of calculations for the biexciton in ZnO/ZnMgO
and CdSe/ZnS NWs. Using Eqs. (\ref{Biexciton Energy1}) and (\ref{Biexciton
Energy2}) the dependence of biexciton energies on interhole distance for
different\ NW radii are calculated and presented in Fig. \ref{Fig5}. The
lower branches of curves correspond to the bonding symmetric state, while
the upper branches of curves correspond to the antibonding antisymmetric
state. In the antisymmetric state when the holes approach one another the
energy increases monotonically for all considered radii of the NW. It
follows from Fig. \ref{Fig5} for the symmetric state when the holes approach
one another the energies are lowered down to a particular distance $R$ after
which the energies rise steeply for when the distance is further decreased.
The minimums of energy for the symmetric state are sharply pronounced and
lie below the energy of two separated excitons for all considered NW radii.
The minimums are much deeper for ZnO/ZnMgO NW and occur when $R>1.8$ nm,
while in the case of CdSe/ZnS the minimums appear at $2<R<3$ nm. The
confinement potentials for the electrons and holes in ZnO/ZnMgO and CdSe/ZnS
NWs are significantly different. The comparison of Fig. \ref{Fig3}a and \ref%
{Fig3}b and \ref{Fig5}a and \ref{Fig5}b illustrates the effect of the
confinement on the energy dependence on interhole distance for trions and
biexciton, respectively. The dependences of trion and biexciton energies on
interhole distance for CdSe/CdS are different compared to CdSe/CdS due to
the smaller confinement potentials and have similarity to that for ZnO/ZnMgO
NW.

\begin{figure}[t]
\begin{center}
\includegraphics[width=8.75cm]{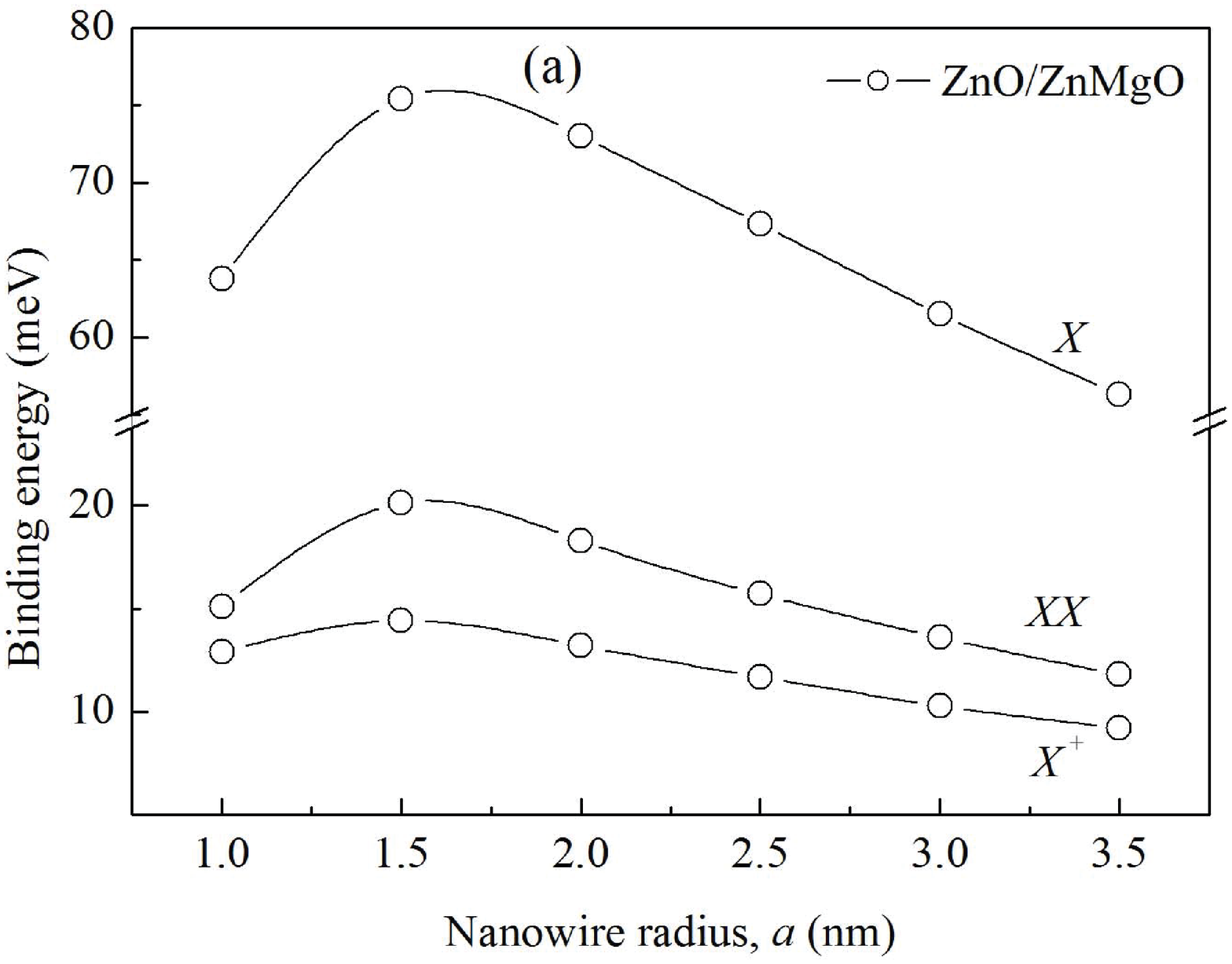} \vspace{-0.1cm} %
\includegraphics[width=8.75cm]{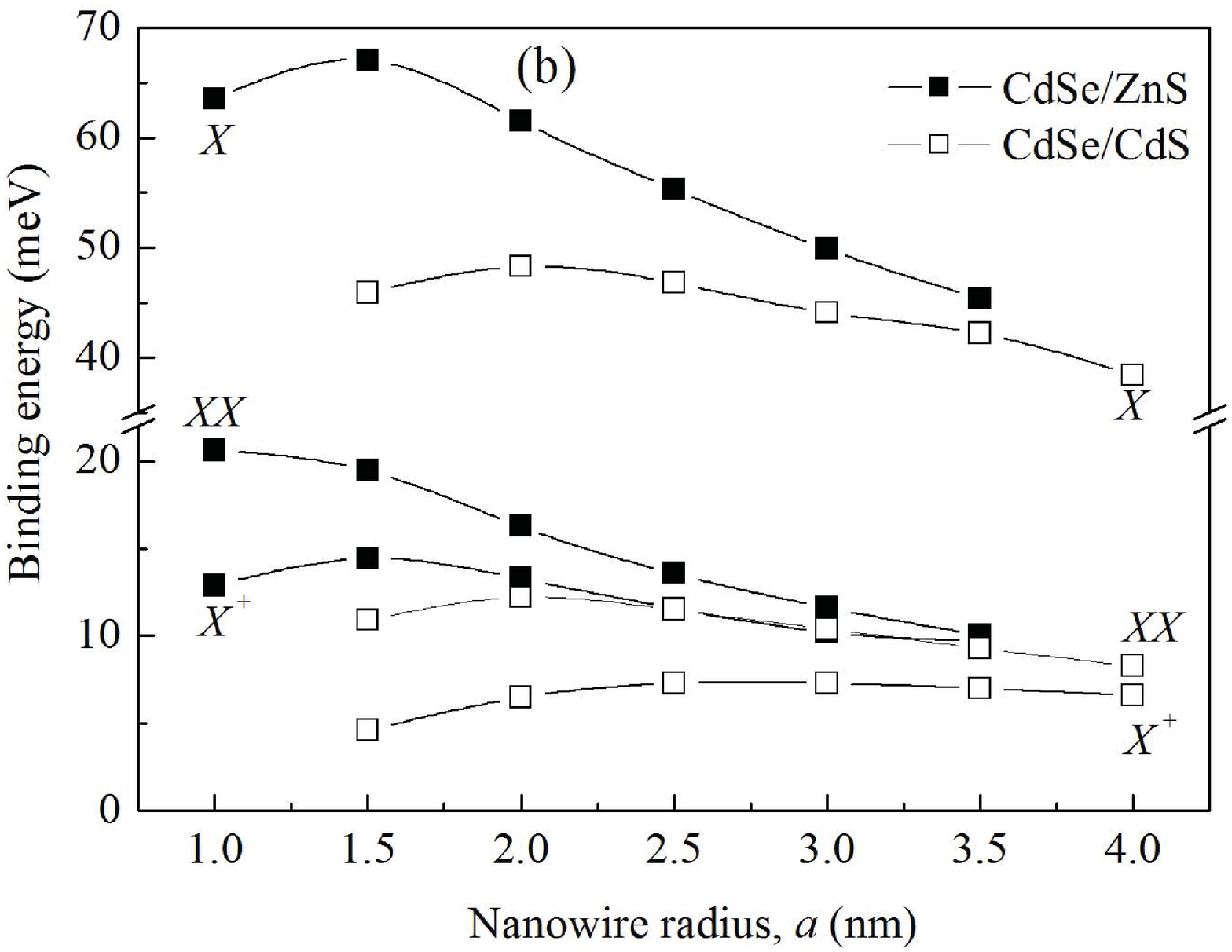} \vspace{-0.1cm}
\end{center}
\caption{ Dependence of the binding energies of excitonic complexes on the
radius of a nanowire in ZnO/ZnMgO ($\circ $) (a), CdSe/ZnS ($\blacksquare $)
and CdSe/CdS ($\square $ ) (b) NWs, respectively, The thickness of the
surrounding dielectric shell is $b=2$ nm. }
\label{Fig1011}
\end{figure}

The results of our calculations demonstrate an appreciable dependence of the
exciton, trion and biexciton binding energy on the radius of NW. In Fig. \ref%
{Fig1011} are presented dependence of the binding energies of the exciton,
trion and biexciton in ZnO/ZnMgO, CdSe/ZnS and CdSe/CdS on the radius of NW.
We perform calculations for the lateral confinement potentials $U_{e}^{0}$
and $U_{h}^{0}$ for a conduction and valence band offset between core and
shell materials presented in Table 1. The comparison of the binding energies
for the exciton, trion and biexciton shows that for the same hole to
electron mass ratio the binding energy of biexciton is larger than for the
trion and smaller than for the exciton in all NWs. Moreover, for the same
input parameters the exciton, trion and biexciton have the maximum binding
energy in ZnO/ZnMgO and CdSe/ZnS NW for the same radius $a=1.5$ nm. In
CdSe/CdS for the set of the lateral confinement potentials $U_{e}^{0}=0.30$
meV and $U_{h}^{0}$ =0.44 meV the maximum of the binding energies for these
excitonic complexes are pronounced at $a=2.0$ nm for the exciton and
biexciton, while the trion has the maximum binding energy for about $80\%$
larger radius. The increase of the confinement potentials in CdSe/ZnS
qualitatively and quantitatively changes the binding energy of the excitonic
complexes: i) the binding energies of excitonic complexes are increased; ii)
they are bound for a smaller NW radius, $a=1.0$ nm; iii) while the biexciton
binding energy is monotonically decreases, the maximum of binding energy for
the exciton and trion occurs at the radii of NW $a=1.5$ nm and $a=2.0$ nm,
respectively. Our calculations show that the biexciton binding energy
exceeds that for the trion rather significantly at a small NWs radius ($%
a=1.5 $ nm in ZnO/ZnMgO and CdSe/CdS, and $a=1.0$ nm in CdSe/ZnS) and with
the increase of the radius the difference between binding energies
decreases. Moreover, in both CdSe/ZnS and CdSe/CdS NWs the
biexciton-to-trion binding energy ratio is greater than unity, decreasing
with the NW radius increase. The same tendency one can observe for in
ZnO/ZnMgO NW when the radius $a\geq 1.5$ nm.

\begin{figure}[t]
\begin{center}
\includegraphics[width=5.8cm]{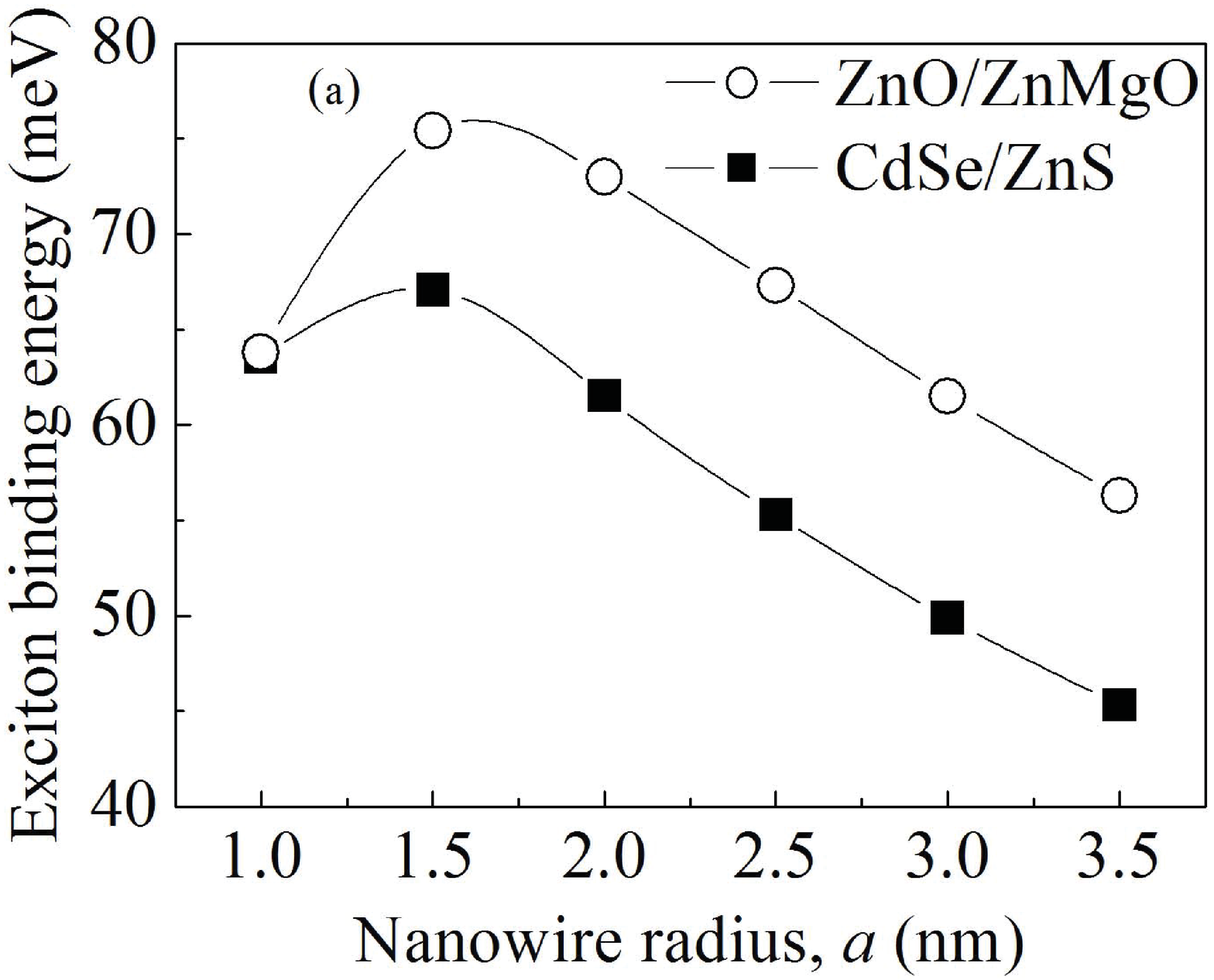} \vspace{-0.1cm} %
\includegraphics[width=5.8cm]{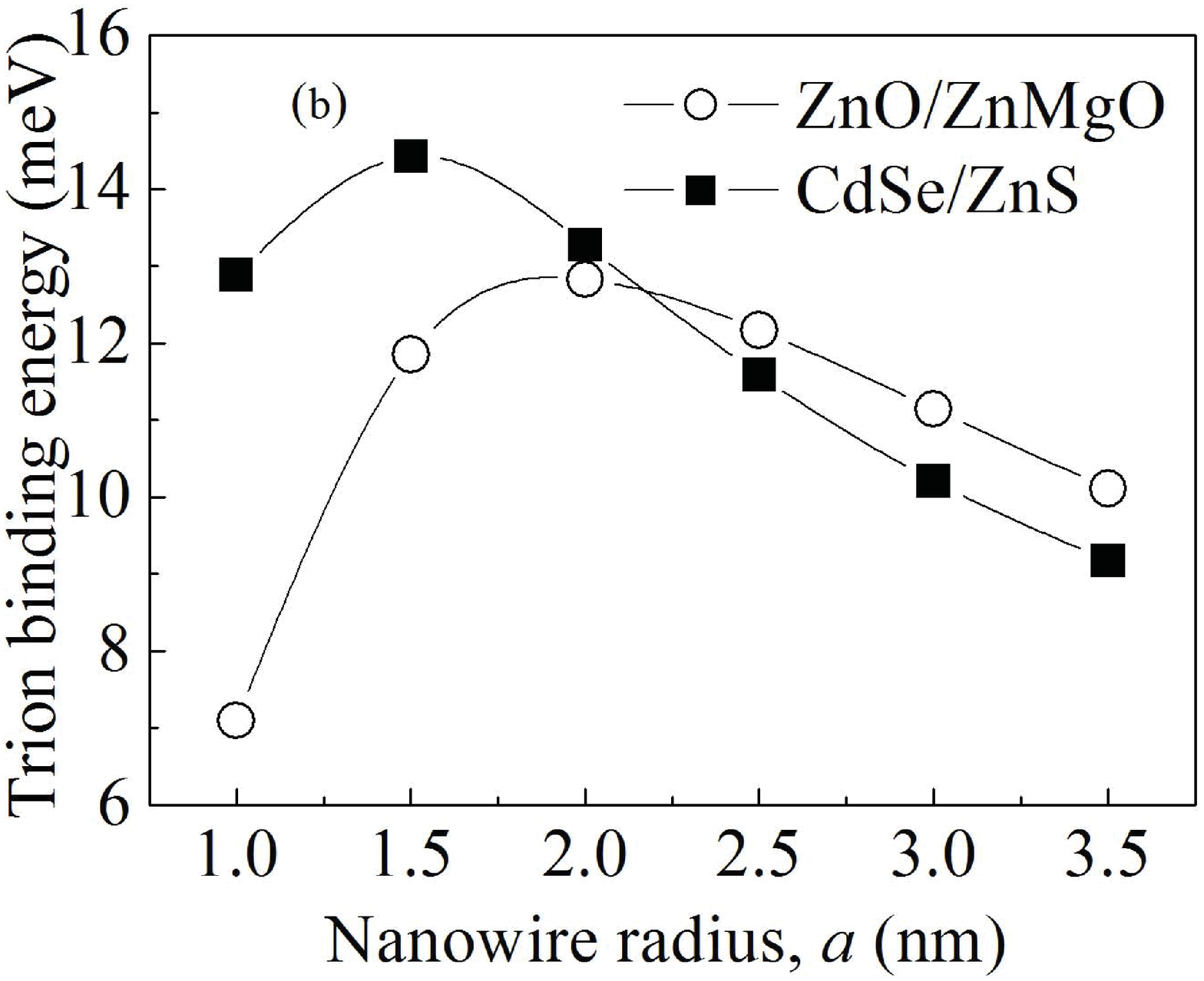} \vspace{-0.1cm} %
\includegraphics[width=5.8cm]{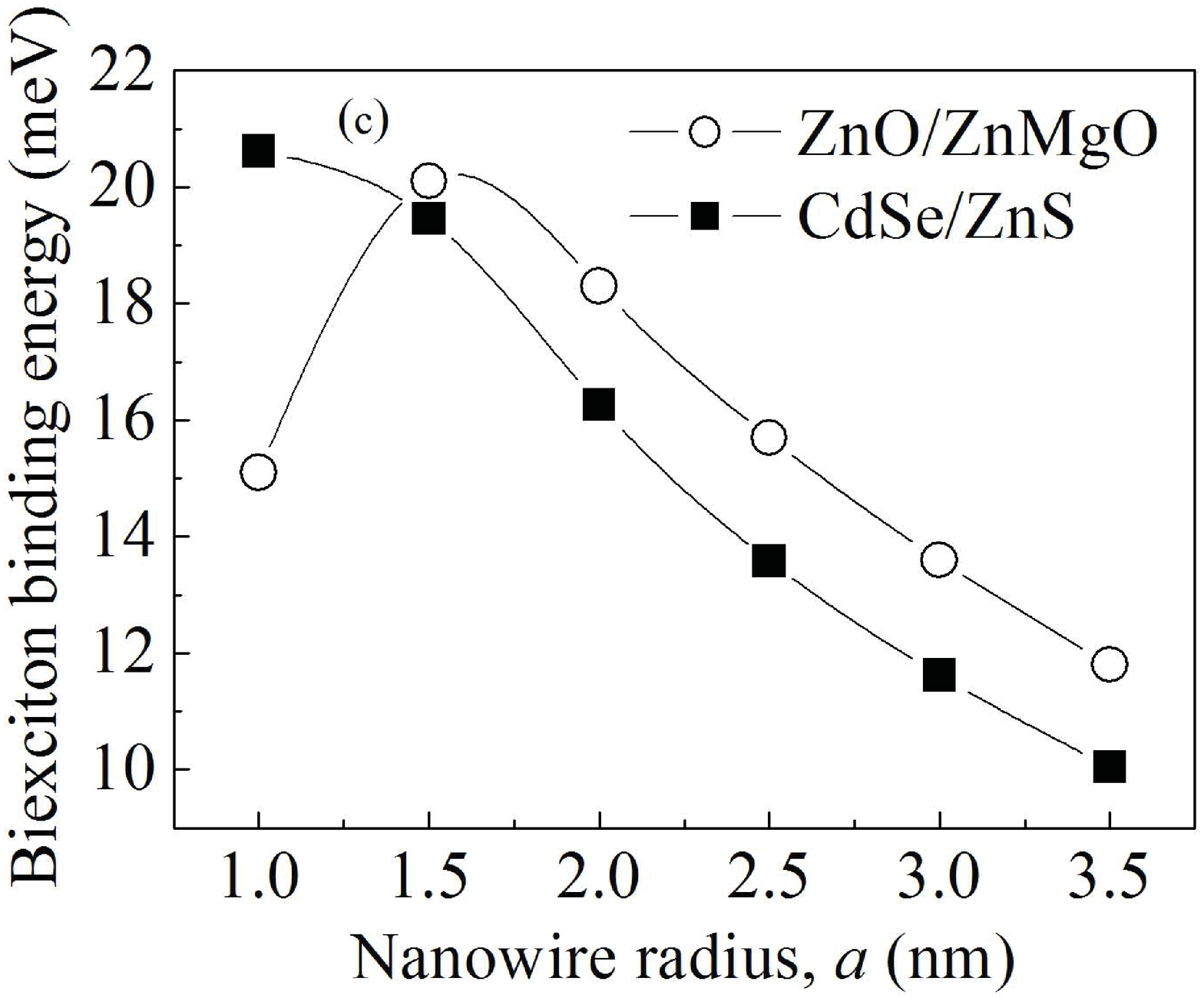} \vspace{-0.1cm}
\end{center}
\caption{Comparison of the dependence of the binding energies of the exciton
(a), trion (b) and biexciton (c) in ZnO/ZnMgO ($\circ $) and CdSe/ZnS ($%
\blacksquare $) NWs on radius of a NW. The thickness of the surrounding
dielectric shell is $b=2$ nm.}
\label{Fig2}
\end{figure}

\begin{figure}[t]
\begin{center}
\includegraphics[width=8.75cm]{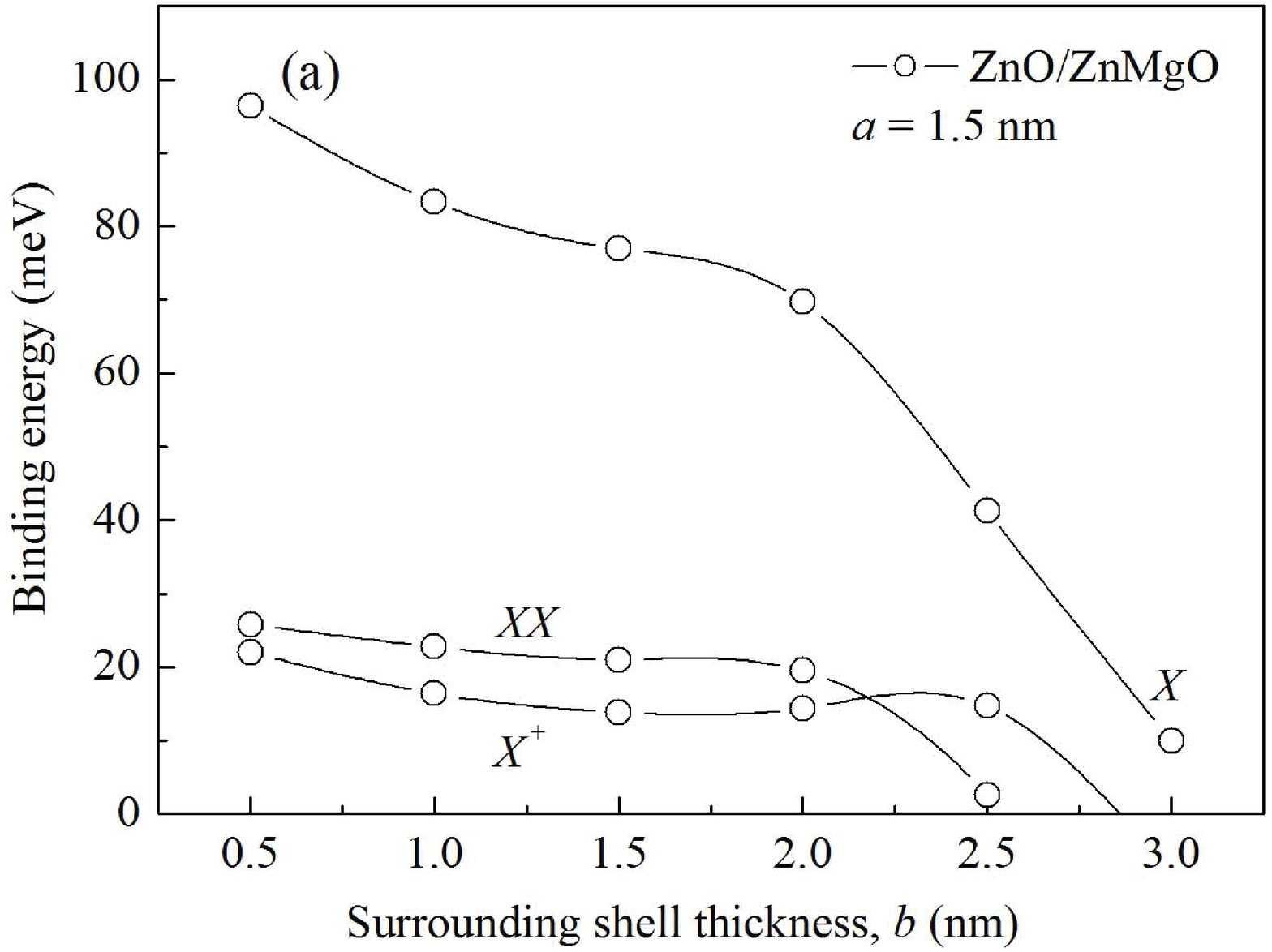} \vspace{-0.1cm} %
\includegraphics[width=8.75cm]{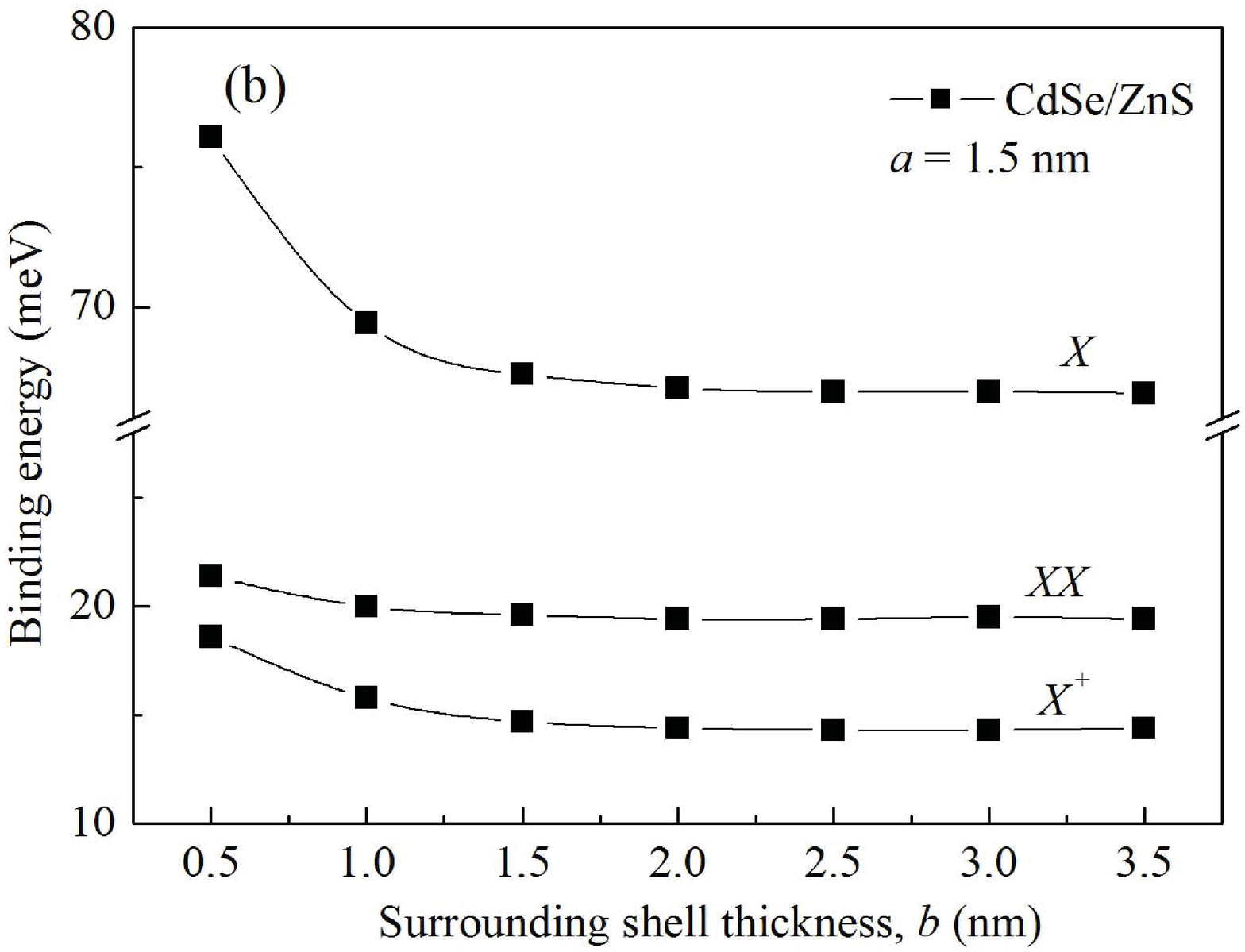} \vspace{-0.1cm}
\end{center}
\caption{Dependence of the exciton, trion and biexciton binding energy in
ZnO/ZnMgO (a) and CdSe/ZnS (b) on the thickness of the surrounding
dielectric shell.}
\label{Fig1214}
\end{figure}

Now let us present, discuss and compare the results for the binding energies
of the exciton, trion and biexciton. In Fig. \ref{Fig2}a the single exciton
binding energies in ZnO/ZnMgO and CdSe/ZnS calculated by means of Eqs. (\ref%
{ExcitEner}) and (\ref{Notation}) are presented. The binding energy varies
between 56.3$-$75.3 meV when the NW radius varies between 1.0$-$3.5 nm, and
has maximum at about 1.5 nm NW radius and at $a=3.5$ nm the exciton binding
energy drops to that of bulk materials. The same behavior of exciton binding
energy is observed for ZnO/ZnMgO quantum wells \cite{Sun2002}, with maximum
at about 2 nm QW width. The dependence of trion binding energy on the NW
radius is depicted in Fig. \ref{Fig2}b. The maximum of the binding energy
12.8 meV is obtained for the NW radius $a=1.5$ nm. With the further decrease
of wire radius the binding energy sharply drops. With increasing of the wire
radius from 2.0 nm to 3.5 nm the binding energy decreases again.

The dependences of the binding energy of biexciton in ZnO/ZnMgO and CdSe/ZnS
on the NW radius is plotted in Fig. \ref{Fig2}c. The biexciton has the
maximum binding energy 21 meV for the wire radius $a=1.5$ nm. Interestingly,
in Ref \cite{Chia} biexcitons were investigated in the ZnO/ZnMgO/ZnMgO
multiple quantum wells. Experimentally determined binding energies for
biexcitons vary between 17.5 to 30.9 meV\ depending on the width of quantum
well and the maximum value was obtained for 2 nm width and binding energy
drops to that of bulk materials.

We also study the influence of the thickness of dielectric shell on the
binding energy of excitonic complexes. In Fig. \ref{Fig1214} are presented
the dependence of the binding energy of the exciton, trion and biexciton
binding energy on the thickness of the surrounding dielectric shell in
ZnO/ZnMgO and CdSe/ZnS. We have calculated the binding energy as a function
of barrier width $b$ for the exciton, trion and biexciton in ZnO/ZnMgO and
CdSe/ZnS NWs. One can conclude that in the case of CdSe/ZnS NW all excitonic
complexes remain stable with the increase of dielectric shell thickness,
while in ZnO/ZnMgO NW biexcitons become unstable when the surrounding
dielectric shell exceeds 2 nm. The trion remains stable for the thickness of
the dielectric shell $b<2.5$ nm. The stability of excitonic complexes in
CdSe/ZnS NW can be explained by the high lateral confinement potentials for
the electron and hole. The relatively low potential barriers for the lateral
confinement of electrons and holes in ZnO/ZnMgO NW allow the penetration of
the corresponding electronic wave functions in the surrounding dielectric
shell area which leads to the strong decrease of the binding energies of the
exciton and biexciton when $b>2$ nm. Our calculations demonstrate that the
size of the core of the NW has stronger influence on the binding energy of
trions and biexcitons compared to the thickness of interfacial alloying. Let
us mention that observation for CdSe/CdS \cite{Nanoscale2014} suggested that
the size of the surrounding shell has equal or less influence on Auger
suppression compared to the radius of the core of NW.

\begin{figure}[t]
\includegraphics[width=13.0cm]{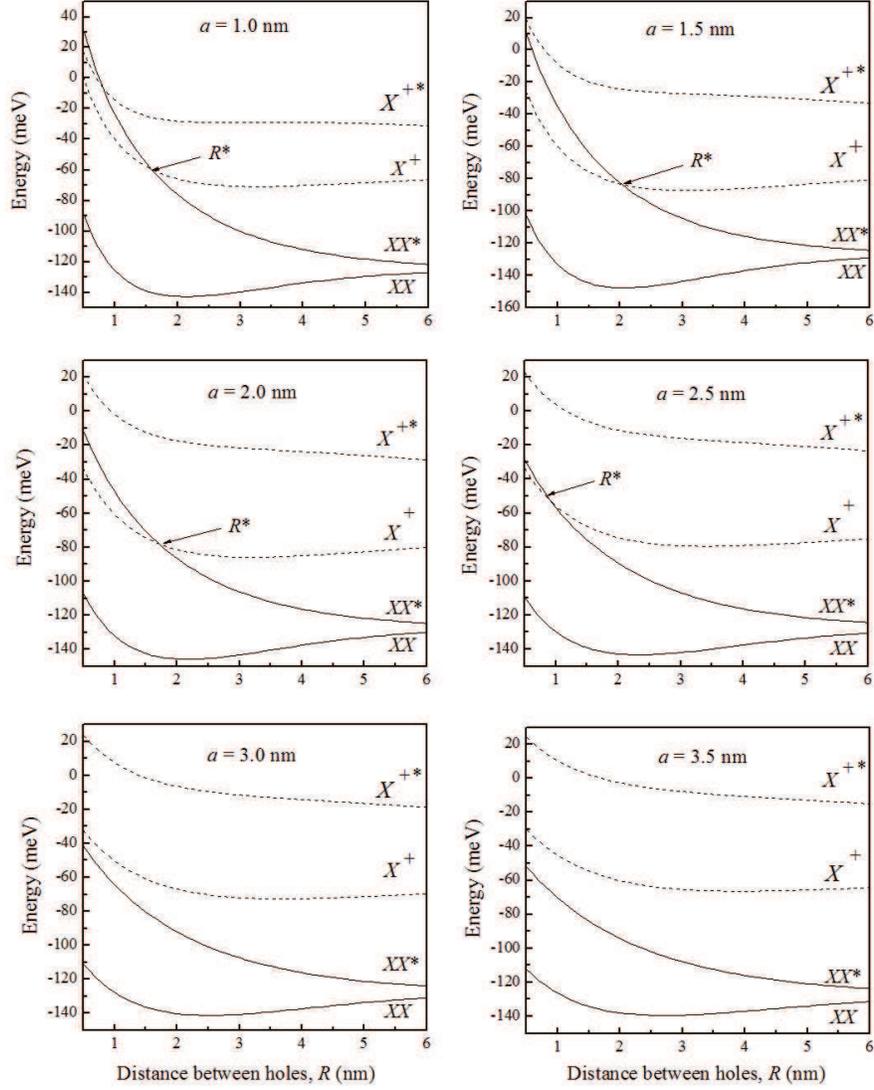} \vspace{-4.5cm}
\caption{ The dependence of the trion and biexciton energies on the
interhole separation for NWs of 1.0$-$3.5 nm radius. The notations $X^{+}$
and $XX$ indicate bounding states energy for the trion and biexciton,
respectively, while $X^{+^{\ast }}$ and $XX^{\ast }$ represents trions and
biexciton antibonding states. $R^{\ast }$ indicates the interhole distance
of the crossing of the biexciton antibonding state energy curve with the
trion bonding state energy curve. Results for ZnO/ZnMgO NW.}
\label{Fig7}
\end{figure}

\begin{figure}[t]
\includegraphics[width=13.0cm]{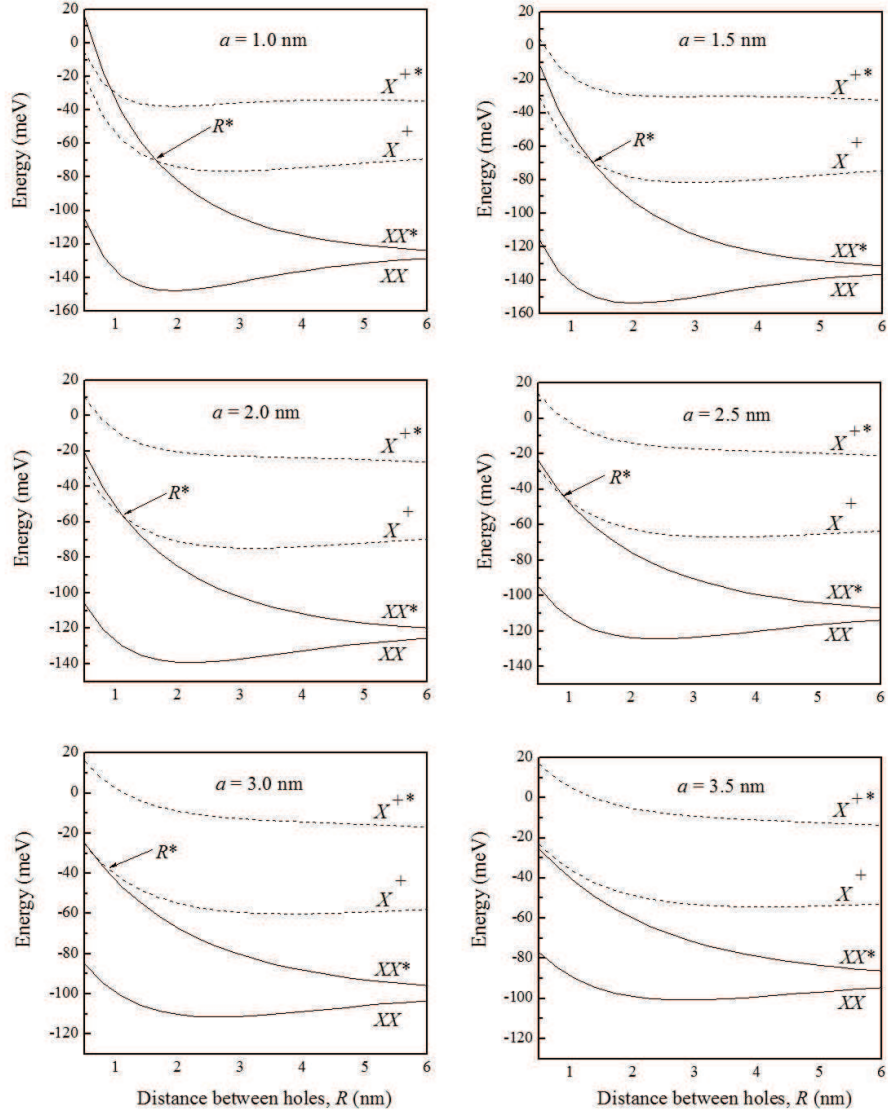} \vspace{-2.0cm}
\caption{ The dependence of the trion and biexciton energies on the
interhole separation for NWs of 1.0$-$3.5 nm radius. The notations $X^{+}$
and $XX$ indicate bounding states energy for the trion and biexciton,
respectively, while $X^{+^{\ast }}$ and $XX^{\ast }$ represents trions and
biexciton antibonding states. $R^{\ast }$ indicates the interhole distance
of the crossing of the biexciton antibonding state energy curve with the
trion bonding state energy curve. Results for CdSe/ZnS NW.}
\label{Fig13}
\end{figure}

Our calculations show a radius dependence of characteristics of excitons,
trions and biexcitons and enhancement of their binding with size reduction
as expected in nanostructures. With the radius reduction quantitative change
of energy level alignment is also observed as one can see in Fig. \ref{Fig7}%
. In Fig. \ref{Fig7} biexciton and trion energies for the same radius of
wire are depicted in the same figure to investigate associative ionization
(AI), the process in which two excitons interact to produce a free electron
and a bound trion. Indeed, a biexciton in the antibonding state can
dissociate into a trion bonding state and an electron. When two excitons
approach one another along Born-Oppenheimer potential energy curve $E(R)$
the ionization occurs only after this reactant pair enters a region of the $%
(E,R)$ plane in which the bound initial electronic state becomes embedded in
the continuum associated with the final state, trion-electron. One can see
that for the radii $a=1.0$ nm, $a=1.5$ nm, $a=2.0$ nm and $a=2.5$ nm of wire
the energy of the antibonding biexciton state sharply increases and the
energy curve crosses the trion bonding state energy curve at the interhole
distance $R^{\ast }$. At distances $R<R^{\ast }$ the biexciton energy in the
antibonding state becomes larger compared to the trion energy. Therefore, it
is possible to transition from the biexciton antibonding state to the trion
bonding state with release of an electron - \textit{e.a.}, the associative
ionization. During photoexcitation the antibonding states of biexciton are
created at different interhole distances $R$. Such states will survive if
holes go away from each other on the distance larger than $R^{\ast }$ before
AI, and then stabilize (dissociate). Quantitatively the probability of
surviving can be estimated by means of $R^{\ast }$. The larger $R^{\ast }$,
the smaller the probability would be. Fig. \ref{Fig7} shows that the
probability should drop with the NW size reduction. In Fig. \ref{Fig13} are
presented the crossings of energy curves for the biexciton in the
antibonding states and trion in the bonding states energies for the same
radii for CdSe/ZnS NW. These results also demonstrate a possible transition
from the biexciton antibonding state to the trion bonding state with release
of an electron at some particular interhole distances $R^{\ast }$. 

Taking into account the behavior of the biexciton binding energy with the NW
size variation, we can propose that there exists an optimal radius of
elongated ZnO/ZnMgO quantum wire, for which biexciton binding energy is
still larger than the bulk value, whilst associative ionization into trion
state (which in its turn has strong tendency to the Auger decay) is
weakened. This radius ranges between $1.5-2.0$ nm. At the same time for the
elongated CdSe/ZnS quantum wire this range is $2.0-2.5$ nm due to the
stronger lateral confinement.

\section{Conclusions}

\label{Conclusions} In summary, we presented the theoretical description of
the trion and biexciton in a NW in the framework of the effective-mass model
using Born-Oppenheimer approximation and considered both the lateral
confinement and the localization potential. The analytical expressions for
the binding energy and wavefunctions are obtained and expressed by means of
matrix elements of the effective one-dimensional cusp-type Coulomb
potentials which parameters are determined self-consistently by employing
the same eigenfunctions of the confined electron and hole states. We
investigated biexcitons and trions in ZnO/ZnMgO, CdSe/ZnS and CdSe/CdS
quantum NWs of a cylindrical shape and study the dependence of their binding
energies on the radius of the NW. It is found that for the same input
parameters the biexciton binding energy in NWs is always larger than binding
energy of the trion. For the same input parameters the exciton, trion and
biexciton have the maximum binding energy for the same radius of ZnO/ZnMgO
NW, while the trion has the maximum binding energy for about $70\%$ larger
radius of a NW. We found an appreciable dependence of the trion binding
energy on the radius of the quantum wire. It was revealed that a radius
reduction up to 1.5 nm enhances binding energy of the exciton, trions and
biexciton in ZnO/ZnMgO NW, while for the biexciton in CdSe/CdS quantum NW
the maximum binding energy is obtained for the thinner NW with 1 nm radius.
For very thin NWs binding energies of excitonic complexes decrease. The
excitonic complexes remain stable in CdSe/ZnS NW with the increase of the
dielectric shell, while in ZnO/ZnMgO NW the trion and biexciton become
unstable when the surrounding dielectric shell exceeds 2.5 nm and 2 nm,
respectively. We suggest \ the mechanism of formation of the trion via
associative ionization of a biexciton. As for probability of the associative
ionization of biexciton into vulnerable to Auger decay trion states, it
continually decreases with increasing the radius of NW. This leads us to the
conclusion that $1-2$ nm radius of NW should be optimal for optoelectronic
application at high excitation intensity.

\appendix

\section{Effective interactions}

\label{app:A} 
\begin{figure}[t]
\begin{center}
\includegraphics[width=8.75cm]{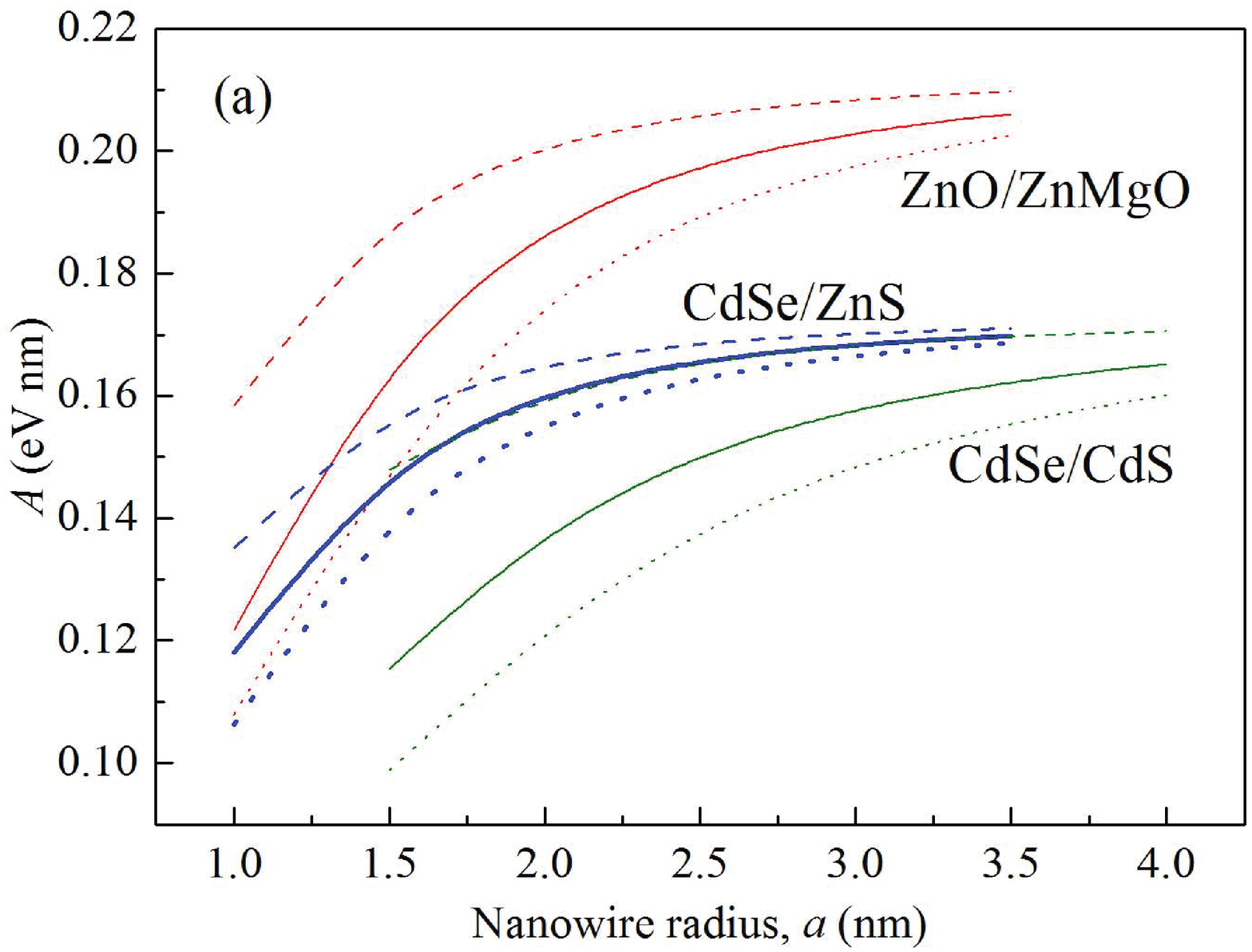} \vspace{-0.1cm} %
\includegraphics[width=8.75cm]{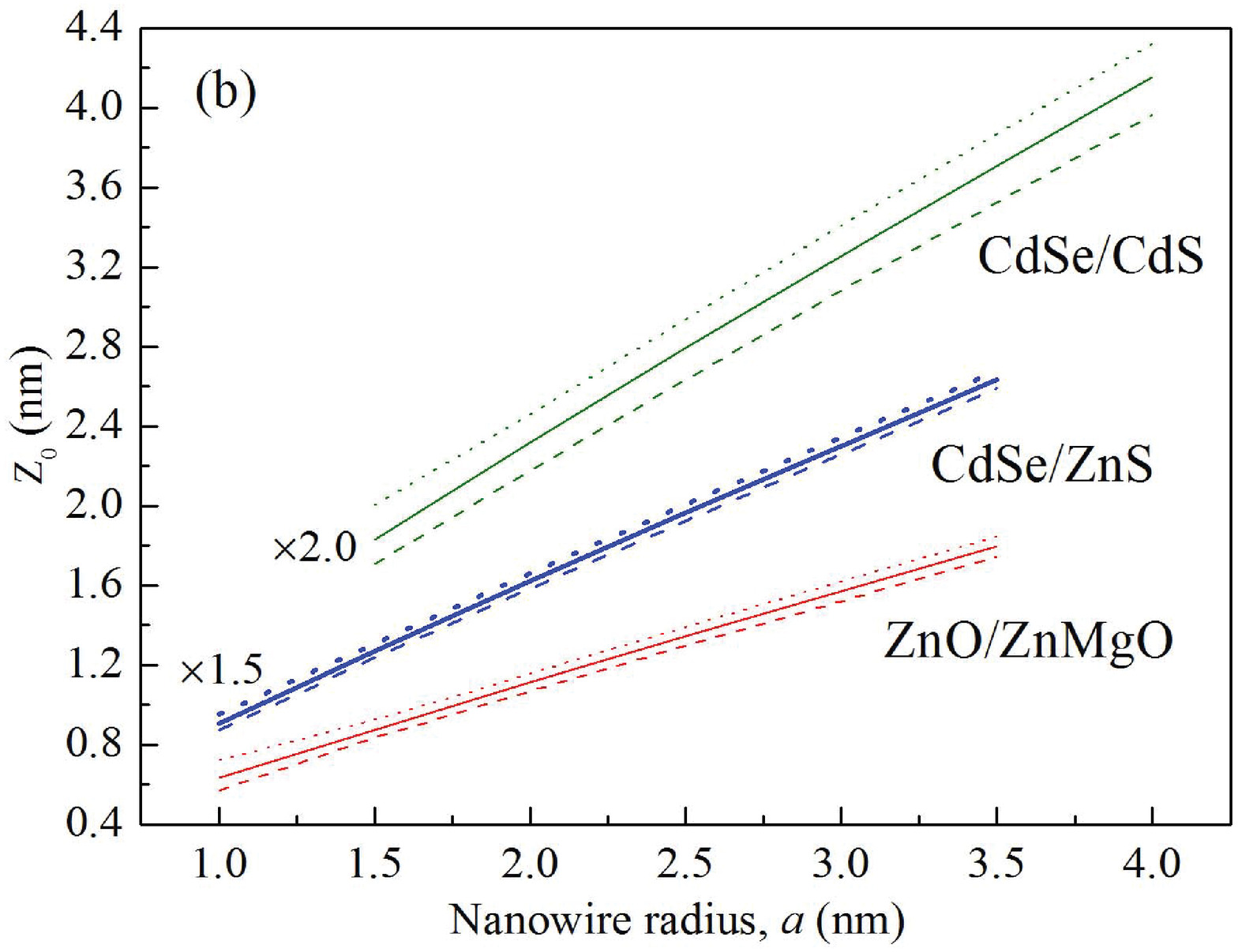} \vspace{-0.1cm}
\end{center}
\caption{ (Color online) Dependence of the fitting parameters $A_{ee},$ $%
A_{eh}$ and $A_{hh}$ (a) and $Z_{0ee},$ $Z_{0eh}$ and $Z_{0hh}$ (b) on the
core radius for ZnO/ZnMgO, CdSe/ZnS and CdSe/CdS NWs. The shell thickness $%
b=2$ nm. Dotted curves - for electron-electron interaction; solid curves -
for electron-hole interaction; dashed curves - for hole-hole interaction.}
\label{Fig89}
\end{figure}

The effective electron$-$hole, hole$-$hole and electron$-$%
electron interactions that are defined as

\begin{eqnarray}
V_{eh}^{eff}(z_{e}-z_{h}) &=&\int\limits_{0}^{a+b}\int\limits_{0}^{2\pi
}\varrho _{e}d\varrho _{e}d\varphi
_{e}\int\limits_{0}^{a+b}\int\limits_{0}^{2\pi }\varrho _{h}d\varrho
_{h}d\varphi _{h}\left\vert \psi _{e}(\rho _{e},\varphi _{e})\right\vert
^{2}V(\mathbf{r}_{e},\mathbf{r}_{h})\left\vert \psi _{h}(\rho _{h},\varphi
_{h})\right\vert ^{2}  \label{Veh} \\
V_{hh}^{eff}(z_{1h}-z_{2h}) &=&\int\limits_{0}^{a+b}\int\limits_{0}^{2\pi
}\varrho _{1h}d\varrho _{1h}d\varphi
_{1h}\int\limits_{0}^{a+b}\int\limits_{0}^{2\pi }\varrho _{2h}d\varrho
_{2h}d\varphi _{2h}\left\vert \psi _{1h}(\rho _{1h},\varphi
_{1h})\right\vert ^{2}V(\mathbf{r}_{1h},\mathbf{r}_{2h})\left\vert \psi
_{2h}(\rho _{2h},\varphi _{2h})\right\vert ^{2}  \label{Vhh} \\
V_{ee}^{eff}(z_{1e}-z_{2e}) &=&\int\limits_{0}^{a+b}\int\limits_{0}^{2\pi
}\varrho _{1e}d\varrho _{1e}d\varphi
_{1e}\int\limits_{0}^{a+b}\int\limits_{0}^{2\pi }\varrho _{2e}d\varrho
_{2e}d\varphi _{2e}\left\vert \psi _{1e}(\rho _{1e},\varphi
_{1e})\right\vert ^{2}V(\mathbf{r}_{1e},\mathbf{r}_{2e})\left\vert \psi
_{2e}(\rho _{2e},\varphi _{2e})\right\vert ^{2}  \label{Vee}
\end{eqnarray}%
where $V(\mathbf{r}_{1},\mathbf{r})=\frac{e^{2}}{\varepsilon \left\vert 
\mathbf{r}_{1}-\mathbf{r}_{2}\right\vert }$ is the Coulomb potential. As a \
result of averaging of $\frac{e^{2}}{\varepsilon \left\vert \mathbf{r}_{1}-%
\mathbf{r}_{2}\right\vert }$ over the electron $\psi _{e}(\rho _{e},\varphi
_{e})$ and hole $\psi _{h}(\rho _{h},\varphi _{h})$ wave functions in the
lateral directions, the effective potentials (\ref{Veh}) - (\ref{Vee}) are
free from the singularity of the bare Coulomb potential at the origin. The
modeling of the potentials (\ref{Veh}) - (\ref{Vee}) with the first order
rational function $\frac{A}{(z+Z_{0})}$, where $z$ is interparticle distance
in $z$-direction and $A$ and $Z_{0}$ are fitting parameters, provides a
slight modification of the long-range Coulomb potential by a cusp-type
Coulomb potential. The values of the fitting parameters $A$ and $Z_{0}$
depend on a dielectric constant $\varepsilon $ of a NW material,\ a NW core
radius $a$ and a shell thickness $b$. In our particular case related to the
ZnO/ZnMgO, CdSe/ZnS and CdSe/CdS quantum NWs the dependence of the fitting
parameters $A$ and $Z_{0}$ on the NW radius are presented in Fig. \ref{Fig89}%
. From Fig. \ref{Fig89}a one can conclude that the fitting parameters $%
A_{ee},$ $A_{eh}$ and $A_{hh}$ that indicate the strength of the cusp-type
Coulomb electron-electron, and hole-hole interactions converge with the
increase of NW radius. The fitting parameters $Z_{0ee},$ $Z_{0eh}$ and $%
Z_{0hh}$ presented in Fig. \ref{Fig89}b display the linear dependence on NW
radius. The slops of these dependences are almost the same for the same NW,
but there are the significant differences for the slopes for ZnO/ZnMgO,
CdSe/ZnS and CdSe/CdS NW. The variation of $Z_{0ee},$ $Z_{0eh}$ and $Z_{0hh}$
parameters with a nanowire radius in ZnO/ZnMgO, CdSe/ZnS and CdSe/CdS NW can
be explained the following way: these parameters are obtained by means of
averaging of 3D Coulomb potential with in-plane (radial) wave functions,
which describe the lateral confinement of electrons and holes. After
averaging procedure Coulomb potentials depend only on the distance between
carriers. $Z_{0}$ should be a measure of their average lateral separation.
Now it is obvious that $Z_{0}$ must increase i) when barrier is decreased
and/or ii) when carrier effective masses are reduced.

\section{Matrix elements for $J$\ and $K$\ }

\label{app:B} The value for $J$\ and $K$\ are given by the following matrix
elements

\begin{eqnarray}
J &=&\left\langle \Phi _{X_{1}}\right\vert \frac{A_{eh}}{\left\vert
z+R/2\right\vert +Z_{0eh}}\left\vert \Phi _{X_{1}}\right\rangle
=\left\langle \Phi _{X_{2}}\right\vert \frac{A_{eh}}{\left\vert
z-R/2\right\vert +Z_{0eh}}\left\vert \Phi _{X_{2}}\right\rangle ,  \label{J}
\\
K &=&\left\langle \Phi _{X_{2}}\right\vert \frac{A_{eh}}{\left\vert
z+R/2\right\vert +Z_{0eh}}\left\vert \Phi _{X_{1}}\right\rangle
=\left\langle \Phi _{X_{1}}\right\vert \frac{A_{eh}}{\left\vert
z-R/2\right\vert +Z_{0eh}}\left\vert \Phi _{X_{2}}\right\rangle .  \label{K}
\end{eqnarray}%
The matrix elements (\ref{J}) can be treated as the total energy of the
cusp-type Coulomb interaction between the hole located at $z=-R/2$ with the
electron density $e\left\vert \Phi _{X_{1}}\right\vert ^{2}$ or the hole
located at $z=R/2$ with the electron density $e\left\vert \Phi
_{X_{2}}\right\vert ^{2}.$ Numerically, these two matrix elements are equal
to one another. The matrix elements (\ref{K}) correspond to the energy of
the cusp-type Coulomb interaction of the overlap charge density $e\left\vert
\Phi _{X_{1}}\Phi _{X_{2}}\right\vert $ localized around the hole located at 
$z=-R/2$ with the hole. By symmetry, the energy of interaction of the
overlap charge density with the hole located at $z=R/2$ has the same value.
The matrix elements $J$\ and $K$\ both depend on the interhole separation $R$
and are calculated for each fixed value of interhole distance.

\section{Matrix elements for $\mathfrak{J}$\ and $\mathfrak{K}$\ }

\label{app:C} For the biexciton, the values for $J$\ and $K$\ are the same
as for the trion and are given by (\ref{J}) and (\ref{K}), while $\mathfrak{J%
}$ and $\mathfrak{K}$ are the matrix elements as defined below

\begin{eqnarray}
\mathfrak{J} &=&\left. \left\langle \Phi _{X_{1}}(z_{1})\Phi
_{X_{1}}(z_{1})\left\vert \frac{A_{ee}}{z_{12}+Z_{ee}}\right. \right\vert
\Phi _{X_{2}}(z_{2})\Phi _{X_{2}}(z_{2})\right\rangle ,  \label{j} \\
\mathfrak{K} &=&\left. \left\langle \Phi _{X_{1}}(z_{1})\Phi
_{X_{2}}(z_{2})\left\vert \frac{A_{ee}}{z_{12}+Z_{ee}}\right. \right\vert
\Phi _{X_{1}}(z_{2})\Phi _{X_{2}}(z_{1})\right\rangle .  \label{k}
\end{eqnarray}%
The matrix element $\mathfrak{J}$ is the repulsion energy of the charge
density $e\left\vert \Phi _{X_{1}}(z_{1})\right\vert ^{2}$\ of electron 1
localized around the hole located at $z=R/2$ with the charge density $%
e\left\vert \Phi _{X_{1}}(z_{1})\right\vert ^{2}$\ of electron 2 localized
around the hole located at $z=-R/2$, when the repulsion occurs via the
cusp-type Coulomb interaction. The matrix element $\mathfrak{K}$ presents
the repulsion energy between the electrons due to the cusp-type Coulomb
interaction, which is connected with the correlation in the motion of the
electrons arising from the antisymmetrization of the wavefunctions in
accordance with the Pauli principle. The matrix elements $\mathfrak{J}$ and $%
\mathfrak{K}$ are functions of the distance $R$ between the holes.

Once the value of $J$\ and $K,$ and $\mathfrak{J}$ and $\mathfrak{K}$ are
known one can find the value of \ $Q$\ and $P$\ as

\begin{equation}
Q=E_{hh}+2J+\mathfrak{J},\text{ \ }P=SE_{hh}+2SK+\mathfrak{K}  \label{Q P}
\end{equation}%
and obtain the biexciton energy as a function of the interhole distance $R$
using Eqs. (\ref{Biexciton Energy1}) and (\ref{Biexciton Energy2}).

\acknowledgments

The authors are thankful to T. Kereselidze for the useful discussion. This
work is supported by joint research grant from Shota Rustavely Georgian
National Science Foundation (\# 04/04) and Science and Technology center in
Ukraine (\# 6207). R.Ya.K. gratefully acknowledges support from the PSC --
City University of New York Award \# 61188-00 49.

\end{document}